\documentclass[aps,pra,showpacs,amsmath,amssymb,twocolumn,nofootinbib,superscriptaddress]{revtex4-1}

\usepackage[pdftex]{graphicx}
\usepackage{mathrsfs,amsmath,amssymb,amsthm,graphicx,caption,subcaption}
\usepackage{fancyhdr}
\usepackage[colorlinks=true,
            linkcolor=red,
            urlcolor=blue,
            citecolor=blue]{hyperref}

\begin{document}

\bibliographystyle{apsrev}

\title{Adaptive-optics-enabled quantum communication: A technique for daytime space-to-Earth links}

\author{Mark T. Gruneisen}
\affiliation{Air Force Research Laboratory, Directed Energy Directorate, Kirtland AFB, NM, United States}
\email[
Approved for public release; distribution is unlimited. Public Affairs release approval AFRL-2021-1343.]{}

\author{Mark L. Eickhoff}
\affiliation{The Boeing Company, PO Box 5670, Albuquerque NM, United States}

\author{Scott C. Newey}
\affiliation{The Boeing Company, PO Box 5670, Albuquerque NM, United States}

\author{Kurt E. Stoltenberg}
\affiliation{The Boeing Company, PO Box 5670, Albuquerque NM, United States}

\author{Jeffery F. Morris}
\affiliation{The Boeing Company, PO Box 5670, Albuquerque NM, United States}

\author{Michael Bareian}
\affiliation{The Boeing Company, PO Box 5670, Albuquerque NM, United States}

\author{Mark A. Harris}
\affiliation{Leidos, Albuquerque NM, United States}

\author{Denis W. Oesch}
\affiliation{Leidos, Albuquerque NM, United States}

\author{Michael D. Oliker}
\affiliation{Leidos, Albuquerque NM, United States}

\author{Michael B. Flanagan}
\affiliation{Leidos, Albuquerque NM, United States}

\author{Brian T. Kay}
\affiliation{Air Force Research Laboratory, Directed Energy Directorate, Kirtland AFB, NM, United States}

\author{Johnathan D. Schiller}
\affiliation{Air Force Research Laboratory, Directed Energy Directorate, Kirtland AFB, NM, United States}

\author{R. Nicholas Lanning}
\affiliation{Air Force Research Laboratory, Directed Energy Directorate, Kirtland AFB, NM, United States}

\date{\today}

\begin{abstract}
Previous demonstrations of free-space quantum communication in daylight have been touted as significant for the development of global-scale quantum networks. 
Until now, no one has carefully tuned their atmospheric channel to reproduce the daytime sky radiance and slant-path turbulence conditions as they exist between space and Earth. 
In this article we report a quantum communication field experiment under conditions representative of daytime downlinks from space. 
Higher-order adaptive optics increased quantum channel efficiencies far beyond those possible with tip/tilt correction alone while spatial filtering at the diffraction limit rejected optical noise without the need for an ultra-narrow spectral filter. 
High signal-to-noise probabilities and low quantum-bit-error rates were demonstrated over a wide range of channel radiances and turbulence conditions associated with slant-path propagation in daytime. 
The benefits to satellite-based quantum key distribution are quantified and discussed.
\end{abstract}
		

\maketitle
\pagestyle{fancy}
\cfoot{Approved for public release; distribution is unlimited. Public Affairs release approval AFRL-2021-1343.}
\lhead{}
\chead{}
\rhead{\thepage}

\section{Introduction}

A century of research in quantum mechanics, optics, computing, and information is culminating in a second quantum revolution.  
Whereas the first quantum revolution explored the foundations of quantum mechanics, the second is focused on utilizing them to change many aspects of the modern world.  
Perhaps the most compelling and ambitious is the quantum internet; a network of quantum computers, repeaters, and memories linked by classical- and quantum-optical channels \cite{van2014quantum, boone2015entanglement, wehner2018quantum}.  
The grand vision consists of a global-scale network whereby satellite-based nodes communicate over quantum channels with ground-based nodes and quantum computers distributed around the globe.  
This network would enable distributed quantum computation, blind quantum computation, quantum-assisted imaging, and precise timing, to name a few proposed applications. 

Progress toward this vision includes seminal demonstrations of quantum communication (QComm) between a satellite and Earth consisting of quantum key distribution (QKD), entanglement distribution, and quantum teleportation \cite{liao2017satellite, yin2017satellite, ren2017ground}.  
These demonstrations were performed at night to avoid daytime optical noise that would have overwhelmed the quantum signal.  
Closely related demonstrations between aircraft and ground \cite{wang2013direct, nauerth2013air, pugh2017airborne} and quantum-limited communication from space \cite{gunthner2017quantum, takenaka2017satellite} were also performed under benign nighttime conditions.  
A robust network however should provide day/night operation.  Numerous terrestrial demonstrations of “daytime” QComm have been reported, including demonstrations performed over long horizontal distances and extreme attenuation \cite{jacobs1996quantum, buttler2000daylight, hughes2002practical, shan2006free, peloso2009daylight, heim2010atmospheric, garcia2013high, carrasco2014correction, liao2017long, vasylyev2017free, gong2018free, arteaga2019enabling}.
However, the quantum-channel atmospheric and radiance conditions were neither reported nor related to actual daytime slant-path channels rendering the relevancy to daytime satellite QComm unsubstantiated.  
Furthermore, long horizontal channels at ground level can have very different atmospheric and background radiance conditions than those encountered in daytime slant-path propagation.  
Unto itself, distance is not a physical property of the free-space channel that directly affects the performance of a daytime quantum channel.  
Rather, it is loss and noise associated with beam divergence, turbulence-induced wavefront errors, and atmospheric scattering.  Traceability to actual implementations require turbulence and background conditions that simulate those of actual slant-path channels.

\begin{figure*}[t!]
	\includegraphics[width=.76\textwidth]{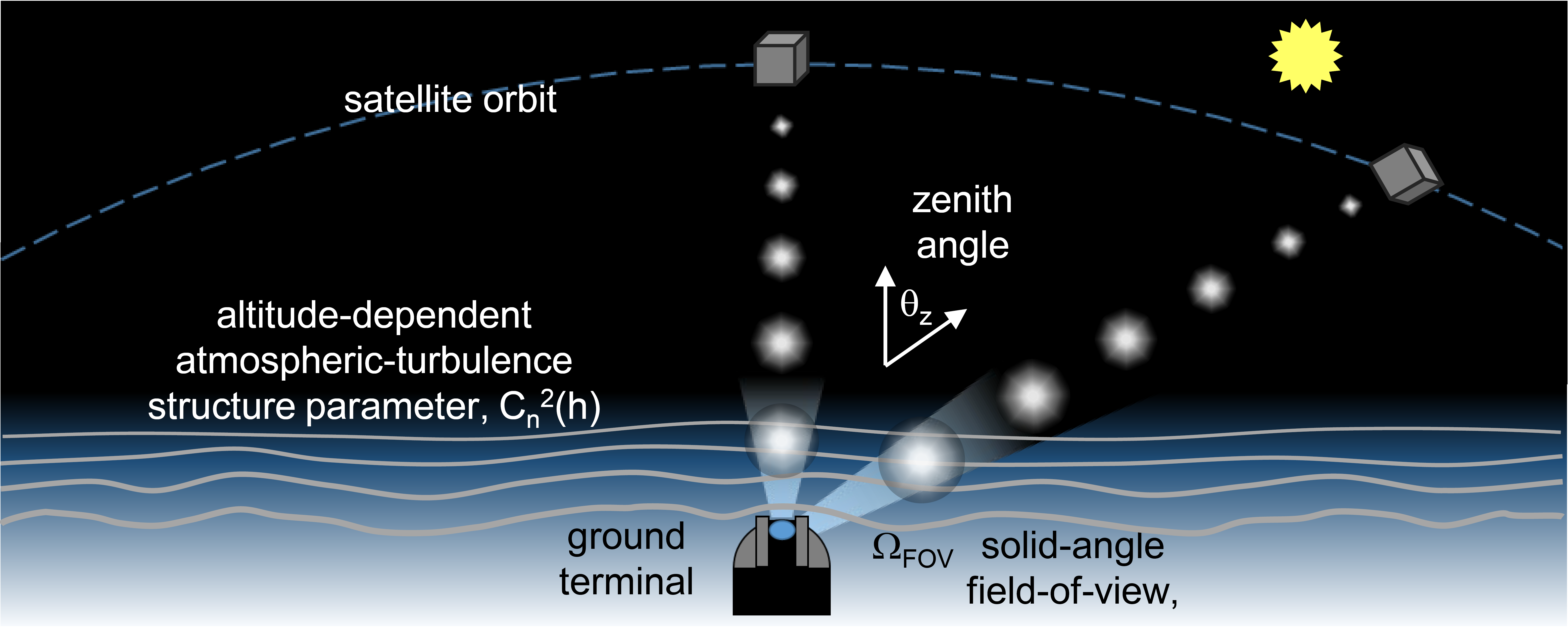}
\caption{\label{fig:downlink_cartoon}
Illustration showing photonic qubits propagating from a satellite through atmospheric turbulence to a ground terminal with sunlight scattering into the telescope's FOV, $\Omega_{\mathrm{FOV}}$. 
	}
\end{figure*}
Figure~\ref{fig:downlink_cartoon} illustrates the downlink scenario in daytime.  
Photonic qubits launched from a satellite propagate through hundreds of km of space and, due to the wave nature of light, expand to many meters in diameter prior to entering the relatively thin atmosphere of Earth.  
As the qubit propagates through the Earth's atmosphere, atmospheric turbulence introduces wavefront errors with the largest contributions occurring near the surface shear layer.  
However, owing to the large beam waist and relatively short turbulence path, additional expansion due to turbulence is a negligible effect.  
The terrestrial ground terminal typically captures only a small portion of the probability amplitude.  
In a satellite pass, aperture-to-aperture coupling losses increase with propagation distance and, depending on the choice of wavelength and aperture sizes, can exceed 10 dB or even 20 dB.  
In daytime, the atmospheric scattering of sunlight into the quantum receiver leads to background photon rates that are proportional to the detector solid-angle field of view (FOV).  
Without optical filtering, background rates can greatly exceed qubit rates.
Optical noise filtering in the temporal and spectral domain is therefore important, but ultimately limited by time-energy uncertainty or the minimum time-bandwidth product of Fourier optics.  
Beyond these limits, further attenuation of the quantum signal is inevitable.  
Consequently, it is important to consider filtering in the spatial domain and the effects of atmospheric turbulence in conjunction with the fundamental limits imposed by diffraction.  

In 2014 and 2016, we presented a concept for integrating AO with a ground terminal for QKD from low-Earth orbit (LEO) \cite{gruneisen2014adaptive, gruneisen2016adaptive}.
The AO system included a fast-steering mirror (FSM) for low-order tip/tilt compensation and a two-dimensional deformable mirror (DM) for higher-order wavefront error compensation.
Through detailed numerical simulations, we showed that a sufficiently high-bandwidth AO system can preserve channel efficiency in the terrestrial receiver while operating at the diffraction-limited (D-L) FOV thereby sharply reducing sky noise and enabling QKD in daylight.  
Turbulence compensation can benefit quantum networks by 1) easing spectral filtering requirements that could hinder protocols based on broadband photons \cite{deng2019quantum}, 2) reducing higher-order spatial modes that compromise momentum indistinguishability that is required for Bell-state measurements \cite{ou2017quantum}, and 3) increasing the efficiency of coupling into single-spatial-mode quantum systems \cite{gruneisen2017modeling, peev2009secoqc, chen2010metropolitan, solano2017dynamics, fatemi2017modal, schneeweiss2013nanofiber, miki2013high, marsili2013detecting, sarovar2016silicon}.

AO is an established technique for real-time sensing and correction of atmospheric-turbulence-induced wavefront errors.  
The concept was proposed in 1953 by Babcock \cite{babcock1953possibility}.  
The first practical design and demonstration was accomplished by Hardy while working for Itek Optical Systems in the mid-1970s \cite{hardy1975real, hardy1977real, duffner2009adaptive}.  
Compensated imaging of satellites in LEO, where high temporal frequencies associated with slewing can challenge AO systems, was first accomplished in 1982 by the Air Force under funding from the Advanced Research Projects Agency \cite{duffner2009adaptive}.  Subsequent advancements made AO an important technology for numerous applications including astronomy, ophthalmology, and microscopy \cite{duffner2009adaptive, fugate1994two, fugate2001handbook, tyson2015principles, roddier1999adaptive}.  
In 2015, one of the authors of this paper implemented AO in a ground terminal to enhance optical communication from the International Space Station by significantly improving single-mode-fiber coupling efficiencies in the presence of atmospheric turbulence \cite{wright2015adaptive}.

In 2017 we extended our satellite-QKD analysis to include the combination of AO and a single-mode optical fiber \cite{gruneisen2017modeling}.  
In this case, the optical mode of the fiber further restricts the telescope FOV, and corresponding background noise, while AO significantly improves the efficiency of qubit coupling into the fiber.  
This analysis shows that a 200-Hz bandwidth closed-loop AO system is sufficient to facilitate daytime QKD even with the high slew rates associated with 400-km and 800-km LEO orbits.
Related analyses explore the potential benefit of AO to satellite QComm by increasing uplink aperture-coupling efficiencies \cite{pugh2020adaptive, oliker2019much}. 

Other attempts to establish relevancy to daytime space-Earth QComm links notably include Liao et al. 2017 which reported a field experiment where a single-mode fiber spatially filtered optical noise \cite{liao2017long}.  
Their architecture however did not expressly simulate a long-distance satellite-to-Earth link.  
Of the 48-dB channel loss reported, only about 2.5 dB was due to non-atmospheric-related aperture coupling as was implied by the reported beam divergence, wavelength and receiver aperture size. 
This is much smaller than the free-space channel loss typically associated with satellite-to-Earth links and roughly equivalent to a satellite with a 20-cm aperture transmitting to a 1-m ground receiver over a distance of only about 110 km.  
About 25 dB was due to low fiber-coupling that is unavoidable without the benefit of AO and low detection efficiencies within the receiver.  
Thus, a majority of the loss was not associated with the free-space channel itself, and therefore it is not clear that these techniques, as they were implemented, are enabling for satellite-to-Earth QComm.  
Furthermore, the large losses in the receiver necessitated the use of an ultra-narrow spectral filter to achieve sufficient signal-to-noise (S/N) probabilities.  
In 2018, Gong attempted an AO implementation but was unsuccessful in compensating atmospheric turbulence due to the slow ($\sim$0.5-s) response time of the AO system \cite{gong2018free}.  
Finally, these experimental demonstrations were reported under ambiguous turbulence and “daytime” background conditions raising additional questions regarding the applicability of either approach to daytime downlinks from space.  
Solving the daytime problem requires that studies be performed under relevant channel conditions and, where AO is concerned, the spatial and temporal characteristics of the turbulent path be understood and accounted for in the design of the AO system.

Recently we demonstrated a pragmatic path forward for realizing robust daytime downlinks for quantum networks \cite{pressrelease, gruneisen2020adaptive}.
Accordingly, we present a field-site validation of our AO-based solution to the daytime sky noise problem.  
Spatial filtering at the diffraction limit allowed the use of a relatively broad 1-nm bandpass spectral filter that is approximately 10, 20, 230, and 1,500 times larger than those used in previous daylight demonstrations \cite{hughes2002practical, liao2017long, shan2006free, jacobs1996quantum}.  
Traceability to daytime slant-path channels was achieved by rigorously characterizing and tuning the atmospheric turbulence and radiance conditions in the channel to match those of daytime slant-path propagation from space.  
We also introduced defocus to create 11 dB of aperture coupling loss representative of diffraction effects over a 700-km propagation distance from space.  
Similar to the downlink scenario, atmospheric turbulence over the horizontal propagation path had a negligible effect on beam divergence.  
Most significantly, we integrated a D-L spatial filter with a 2-kHz frame rate, 130-Hz closed-loop bandwidth AO system.  
The AO system was designed to accommodate the spatial and temporal characteristics of the field-site turbulence as well as those intrinsic to slant-path turbulence.
The experiment was conducted with a stationary transmitter and orbit-dependent slew dynamics were not simulated.  
A supporting analysis that includes LEO slew dynamics shows that the 130-Hz bandwidth system could be sufficient to enable daytime QComm from LEO.
However, slew dynamics that affect AO bandwidth requirements can be considered separately from the intrinsic atmospheric effects.
Higher bandwidth AO systems have already been demonstrated for LEO applications as cited above.  
Results from this field experiment verify that AO enables high S/N detection-probability ratios, low Quantum-Bit-Error Rates (QBERs), and positive QKD bit-yield probabilities in daylight over a wide range of sky angles without the need for ultra-narrow spectral filtering.

Section~\ref{sec:section2} proceeds by reviewing the relationship between optical noise and FOV in a ground terminal for both D-L and turbulence-limited (T-L) FOV scenarios.  
This comparison motivates the concept of operating an optical receiver at the D-L FOV while compensating turbulence-induced wavefront errors with AO.  
Section~\ref{sec:section3} describes our approach to simulating slant-path turbulence at a terrestrial field site.  
Specifically, a 1.6-km horizontal path introduces atmospheric scintillation and spatial coherence commensurate with propagation at zenith angles $\theta_z$ ranging from 0$^{\circ}$ to 76$^{\circ}$.  
A supporting theoretical analysis shows that propagation over longer horizontal distances introduces deep turbulence effects that are not characteristic of canonical slant-path propagation.  
Channel radiance is also tuned to a range from 1 to 100 $\mathrm{W} / ( \mathrm{m}^2 \, \mathrm{sr} \, \mu \mathrm{m})$ commensurate with the daytime sky hemisphere excluding close proximity to the sun angle.  
Turbulence and radiance conditions associated with experimental data sets are correlated to sky angles over the daytime sky hemisphere.  
The temporal characteristics of both the field-site turbulence and the AO system are compared to those encountered in slant-path turbulence with and without the effects of ground telescope slewing.  
Section~\ref{sec:section4} describes the experimental methods employed to integrate the AO system with a qubit prepare-and-measure scheme.  
Section~\ref{sec:section5} presents analyses of the field experiment results.  
The effects of atmospheric scintillation and spatial coherence on system Strehl and quantum channel efficiency are presented both with and without the benefit of higher-order AO.  
The corresponding measured values for S/N probabilities and QBERs are presented as a function of channel radiance.  
Finally, the associated QKD bit yields that could be obtained over these channels are presented and discussed.

\section{Diffraction-limited versus turbulence-limited sky noise filtering}\label{sec:section2}

\begin{figure}[t]
\includegraphics[width=1\columnwidth]{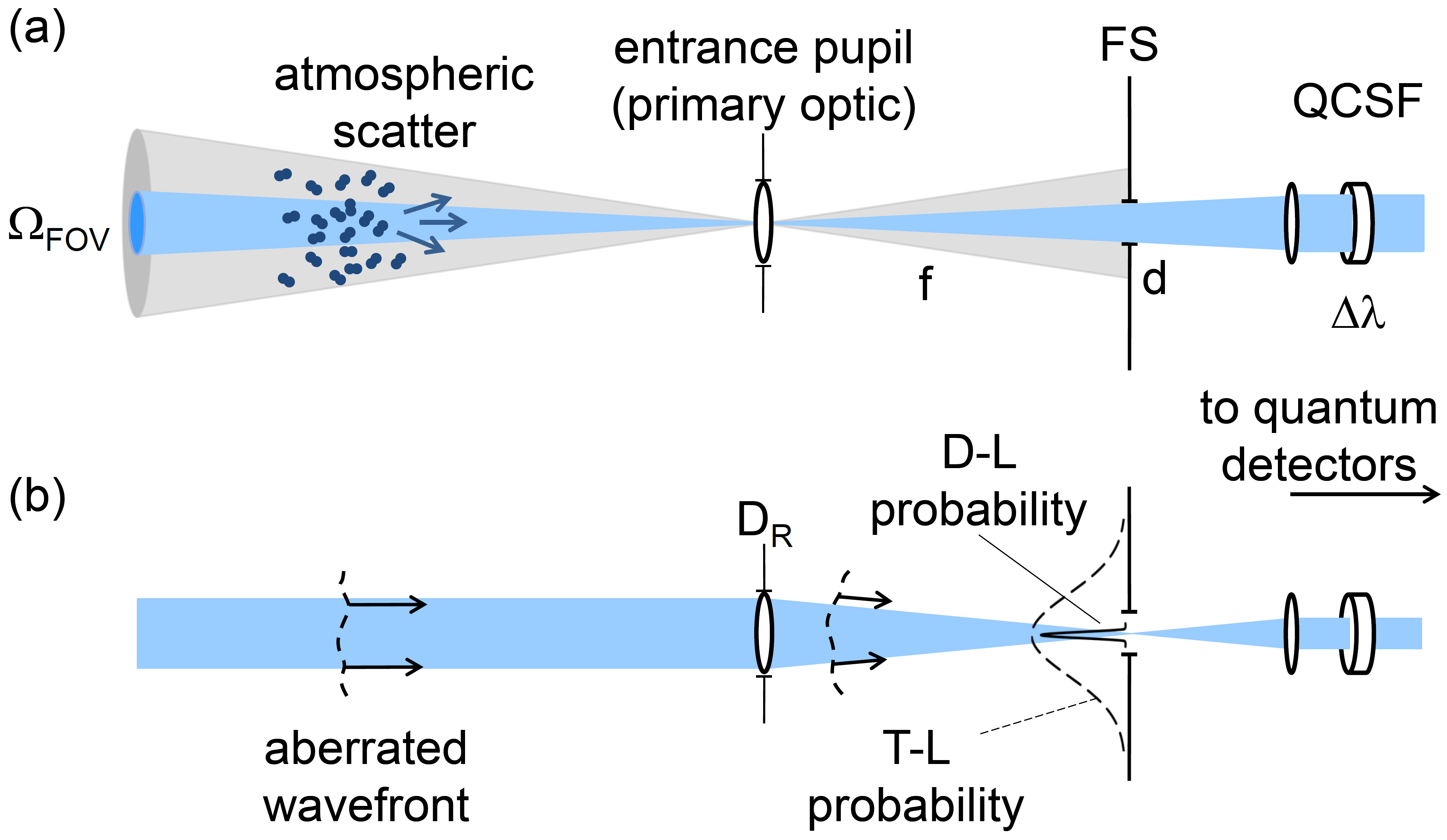}
\caption{\label{fig:receiver_cartoon}
Conceptual schematic of relevant elements in a quantum receiver illustrating (a) the spatial filtering of background noise by the field stop and (b) the effects of turbulence-induced wavefront errors on qubit probability distributions at the field stop.
	}
\end{figure}

Similar to Ref.~\cite{gruneisen2016adaptive}, Fig.~\ref{fig:receiver_cartoon} is a simplified schematic showing pertinent components of a ground receiver telescope.  
Components relevant to this analysis include a receiver primary optic of diameter $D_\mathrm{R}$ followed by a field stop (FS) in the focal plane, a collimating lens, and a quantum-channel spectral filter (QCSF) that transmits light to the quantum detection system. 
Figure~\ref{fig:receiver_cartoon}(a) illustrates the relationship between the FS, the solid-angle FOV $\Omega_{\mathrm{FOV}}$, and the volume of atmospheric scattering that contributes to detector background noise. 
The FS constrains $\Omega_{\mathrm{FOV}}$ and correspondingly the volume of atmospheric scattering that leads to optical noise in the quantum channel.
The number of sky-noise photons, $N_b$, entering the primary optic within a given spectral band and temporal window is proportional to $\Omega_{\mathrm{FOV}}$ as given by Ref.~\cite{er2005background}, 
\begin{equation}\label{eq:NUMBER}
N_b = \dfrac{H_b(\lambda) \, \Omega_{\mathrm{FOV}} \, \pi D^2_{\mathrm{R}} \lambda \, \Delta \lambda \, \Delta t }{4 h c},
\end{equation}
where $H_b (\lambda)$ is the sky radiance in $\mathrm{W} / ( \mathrm{m}^2 \, \mathrm{sr} \, \mu \mathrm{m})$, $\lambda$ is the quantum channel wavelength, $\Delta \lambda$ is the spectral filter bandpass in $\mu$m, $\Delta t$ is the integration time for photon counting, $h$ is Planck's constant, and $c$ is the speed of light. 

Figure~\ref{fig:receiver_cartoon}(b) illustrates the effect of turbulence-induced wavefront errors on the focal-plane distributions.
Wavefront errors enlarge the qubit probability distribution and reduce the probability of transmission to the quantum detectors.   
In a system design, the field stop should be made sufficiently large to minimize qubit losses but otherwise made small to minimize the transmission of scattered sunlight through the FS.  
In the absence of wavefront errors, one strategy for maximizing qubit transmission while reducing background noise is to choose a FS diameter that passes the central lobe of the Airy function associated with a uniform amplitude plane wave brought to focus.  
In this case, a FS of diameter $d_{\mathrm{DL}} = 2.44 \lambda f / D_{\mathrm{R}}$ transmits photons with about 84$\%$ efficiency and the D-L FOV is,
\begin{equation}\label{eq:DLFOV}
\Omega_{\mathrm{DL}} = \pi \Big(  \dfrac{1.22 \lambda}{D_{\mathrm{R}}} \Big)^2 .
\end{equation}
Notice that the D-L FOV increases quadratically with wavelength indicating that spatial filtering is significantly more effective at shorter wavelengths.  
Substituting $\Omega_{\mathrm{DL}}$ for $\Omega_{\mathrm{FOV}}$ in Eq.~\ref{eq:NUMBER}, $N_b$ for the D-L case becomes,
\begin{equation}\label{eq:NUMBERDL}
N_{b \: \mathrm{DL}} = \dfrac{(1.22 \pi)^2 \, H_b(\lambda) \,  \lambda^3 \, \Delta \lambda \, \Delta t }{4 h c}.
\end{equation}
Thus, the number of background photons is now independent of $D_{\mathrm{R}}$ and the explicit wavelength dependence is now cubic.
\textit{D-L spatial filtering allows the ground telescope to be scaled to larger aperture sizes to increase qubit collection efficiency without increasing background noise.}

In the presence of atmospheric turbulence, with aperture sizes that are much larger than the spatial scale of turbulence, higher-order wavefront errors lead to even more broadening of the focal-plane spot than do low-order tip and tilt errors \cite{parenti1992adaptive}.  In this regime, the FS diameter that passes approximately 84$\%$ of incident photons can be approximated by the empirically derived expression $d_{\mathrm{TL}} \approx 2 \lambda \, f / r_0 (\lambda)$, where $r_0 (\lambda)$ is the wavelength-dependent Fried parameter describing the spatial scale of atmospheric turbulence \cite{gruneisen2016adaptive, fried1966optical}.  
The corresponding T-L solid angle is,  
\begin{equation}\label{eq:TLFOV}
\Omega_{\mathrm{TL}} \approx \pi \Big(  \dfrac{\lambda}{r_0 (\lambda)} \Big)^2 .
\end{equation}
For ground telescopes with apertures larger than $r_0 (\lambda)$, the D-L FOV is smaller than the T-L FOV by a factor of,
\begin{equation}\label{eq:FOVRATIO}
\dfrac{\Omega_{\mathrm{DL}}}{\Omega_{\mathrm{TL}}} =  \Big(  \dfrac{1.22 \, r_0 (\lambda)}{D_{\mathrm{R}}} \Big)^2 .
\end{equation}
For the cases modeled in Refs.~\cite{gruneisen2016adaptive, gruneisen2017modeling} with $D_{\mathrm{R}}$=1 m, $\lambda$=780 nm, and $r_0$ ranging from 8.5 cm at zenith to 2.9 cm at $\theta_z = 60^{\circ}$, Eq.~\ref{eq:FOVRATIO} indicates that the D-L FOV decreases optical noise by factors of approximately 100 to 800, respectively, relative to operating at the T-L FOV.  
With perfect turbulence compensation, this reduction in noise could be achieved without introducing signal loss.  

In principle, a well-designed AO system can restore an aberrated wavefront to near-D-L quality.  
In practice, the degree of turbulence compensation achieved will depend on the AO system's ability to spatially and temporally resolve turbulence-induced wavefront errors.  
Designing a relevant field experiment therefore requires understanding the nature of slant-path turbulence, how this can be simulated at a terrestrial field site, and compatibility between atmospheric conditions and various AO design techniques.

\section{Simulating a daytime slant-path channel at a terrestrial field site}\label{sec:section3}

Meaningful simulations of daytime slant-path quantum channels require certain channel conditions be duplicated while others can be introduced at a reduced scale without loss of relevancy.  
For example, Eq.~\ref{eq:NUMBERDL} shows that when operating at the D-L FOV, as was done in this field experiment, $N_b$ is independent of $D_{\mathrm{R}}$ and depends only on $H_b(\lambda)$ and $\lambda$.  
Therefore, it was necessary to introduce a realistic range of channel radiances independent of our choice for $D_{\mathrm{R}}$.  
The effects of turbulence over horizontal paths can be significantly different than those encountered in slant-path propagation.  
It was therefore necessary to choose a path \textit{distance} that was representative of slant-path propagation from space.  
Aperture-to-aperture coupling losses commensurate with diffraction and a 700-km propagation path were introduced over a much shorter distance via defocus.  
The receiver aperture diameter was chosen to be sufficiently large that the benefit of AO to higher-order wavefront compensation could be explored.
Results can be scaled to larger aperture telescope systems by increasing the number of subapertures in the AO system.  
Because the quantum channel wavelength affects a number of quantum channel characteristics, the choice of wavelength is addressed first.

\subsection{Wavelength selection}

Previous analyses have considered both 1550-nm and wavelengths near the 775-nm second harmonic \cite{hughes2002practical, liao2017long, nordholt2002present, bourgoin2013comprehensive}.  
For a receiver operating at the diffraction limit and with a D-L FOV, estimates based on local atmospheric conditions indicate the daytime S/N probabilities and qubit-transmission rates can be better near 775 nm.  
MODTRAN simulations under normal haze conditions indicate that local sky radiances near zenith are on average about 16-times greater near 775 nm. 
However, the two-times higher photon energy and four-times smaller FOV indicated by Eq.~\ref{eq:DLFOV}, result in the number of noise photons being only two-times greater.  
In many cases, the geometrical aperture-to-aperture coupling efficiency can be approximated by the Friis equation \cite{friis1971introduction},
\begin{equation}\label{eq:FRIIS}
\eta_{\mathrm{geo}} = \Big(  \dfrac{\pi \, D_{\mathrm{T}} \, D_{\mathrm{R}}}{4\, \lambda \, z} \Big)^2,
\end{equation}
where $D_{\mathrm{T}}$ is the transmitter diameter and $z$ is the propagation distance.  
In these cases, the aperture coupling efficiency is approximately four-times greater at 775 nm.  
MODTRAN simulations performed under normal haze conditions also indicate that atmospheric transmission near zenith is about 90$\%$ of that at 1550 nm.  
Altogether, this suggests that near the 775-nm wavelength, the S/N probability ratio is about 1.8-times better and the signal photon rate is 3.6-times greater.  
The near-visible wavelength also allows the use of reasonably efficient, compact, low-noise, and inexpensive silicon-based detectors and cameras.  
A more detailed analysis that includes the wavelength-dependent nature of both turbulence-induced wavefront errors and AO compensation shows that quantum channel performance is generally better at shorter wavelengths \cite{lanning2021quantum}. 
Optimal wavelength selection for any space-Earth link should be based on an analysis of site-specific conditions including the effects of scattering, atmospheric turbulence, and the ability to implement AO effectively.  
Accordingly, we chose 780 nm for the quantum channel wavelength.

\subsection{Simulating slant-path turbulence over a terrestrial path}
Optical propagation over long terrestrial paths bears little relevancy to slant-path propagation due to the onset of deep turbulence effects.  
Atmospheric turbulence is initiated by convective air movement and the strength of turbulence decreases rapidly with increasing distance from the surface shear layer.  
This aspect of turbulence has been quantified through altitude-dependent models of turbulence based on the atmospheric structure parameter function, $C_n^2 (h)$, where $h$ is the height above the ground level.  
For many ground observational sites, the Hufnagel-Valley HV$_{5/7}$ model of $C_n^2 (h)$ defines relevant conditions \cite{sasiela2012electromagnetic}.  
To account for varying strengths of turbulence, the model can be scaled through a multiplicative factor to $C_n^2 (h)$.
These scaled versions are designated by the multiplicative factor such that a 2$\times$HV$_{5/7}$ profile describes turbulent layers that are twice as strong as those in the 1$\times$HV$_{5/7}$ profile.  
The net effect of turbulence over any path is found by considering the integrated path.   

The effects of atmospheric turbulence on the spatial characteristics of the optical field are characterized through spatial coherence and scintillation properties.  
Spatial coherence is quantified through Fried's coherence length, $r_0$ \cite{fried1966optical}.  
Scintillation associated with the depth of the turbulent path can be described through the log-intensity variance, or Rytov variance.  
Experimentally, scintillation is quantified via the scintillation index, $\sigma_I^2$, which is a measurable quantity that saturates with increased depth of turbulence \cite{andrews2004field}.  
Appendix~\ref{sec:appendixA} reviews theoretical expressions for $r_0$, Rytov variance, and $\sigma_I^2$ for both horizontal and slant-path propagation.  
From these expressions, one can calculate the horizontal propagation distance that is required to introduce $r_0$ and Rytov variances comparable to those introduced by slant path propagation at various zenith angles:
\begin{equation}\label{eq:LR0}
\begin{split}
L_{r_0} &= \Bigg[  \dfrac{2.64  \int_{0}^{a} C_n^2 (h) \,dh}{C_n^2 (h_0)} \Bigg] \sec(\theta_z) \\
&= [377.1 \,\mathrm{m}] \sec(\theta_z)
\end{split}
\end{equation}
and
\begin{equation}\label{eq:LRYTOV}
\begin{split}
L_{\mathrm{Rytov}} &= \Bigg[  \dfrac{4.5  \int_{0}^{a} C_n^2 (h) \, h^{5/6} \,dh}{C_n^2 (h_0)} \Bigg]^{6/11} \sec(\theta_z) \\
&= [682.1 \,\mathrm{m}] \sec(\theta_z)
\end{split}
\end{equation}
where $a$ is the altitude of the satellite transmitter and the coefficients 377.1 m and 682.1 m are found by numerical integration assuming the height of horizontal propagation is $h_0$=10 m, consistent with our field-site conditions.  
Note that because the overall scaling of turbulence strength (i.e., 1$\times$HV$_{5/7}$ vs 2$\times$HV$_{5/7}$) can occur at any altitude, these equivalent path lengths are independent of turbulence strength.  

\begin{figure}[t]
	\includegraphics[width=1\columnwidth]{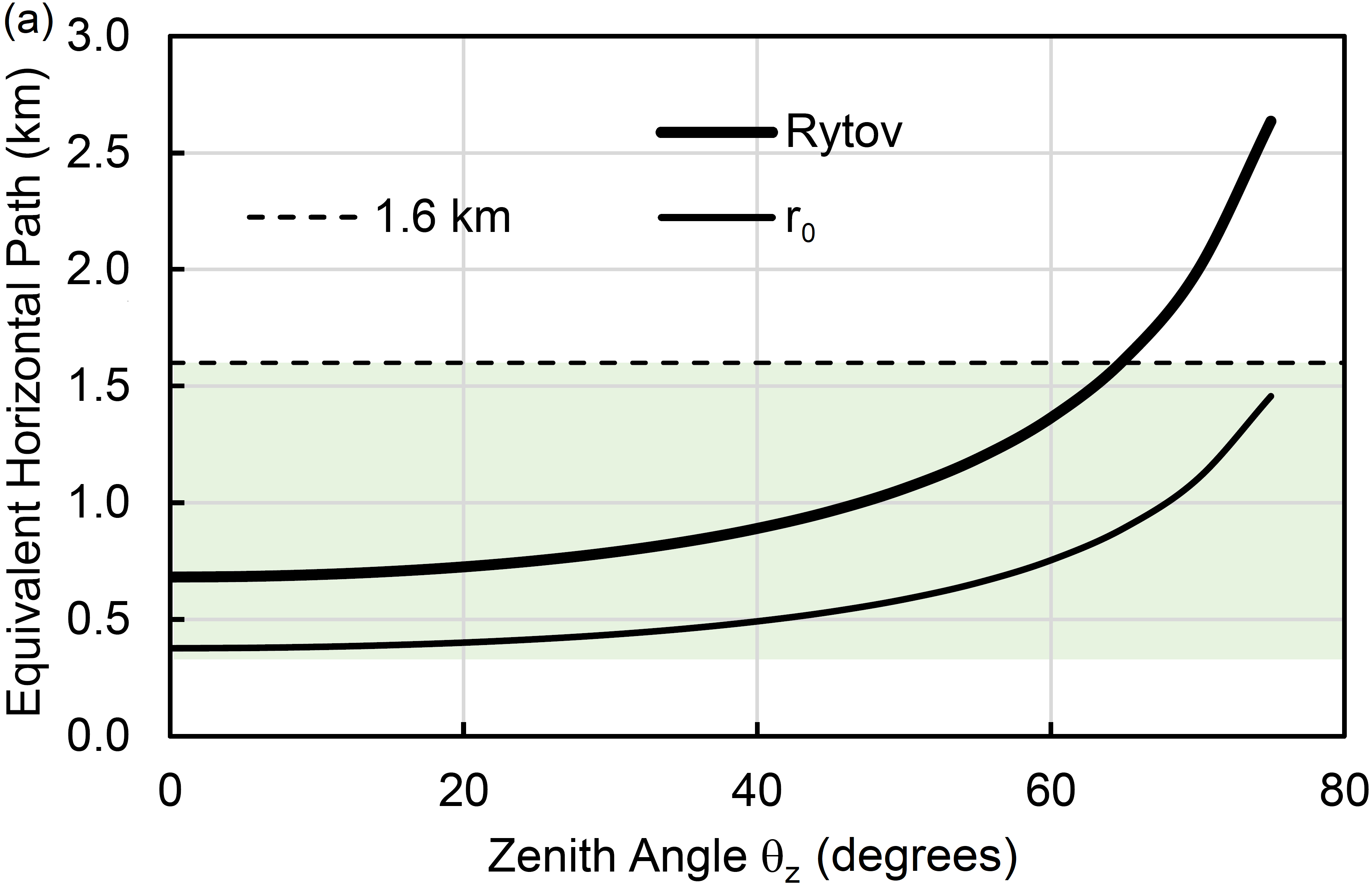}\\
      \vspace{0.25cm} 
	\includegraphics[width=1\columnwidth]{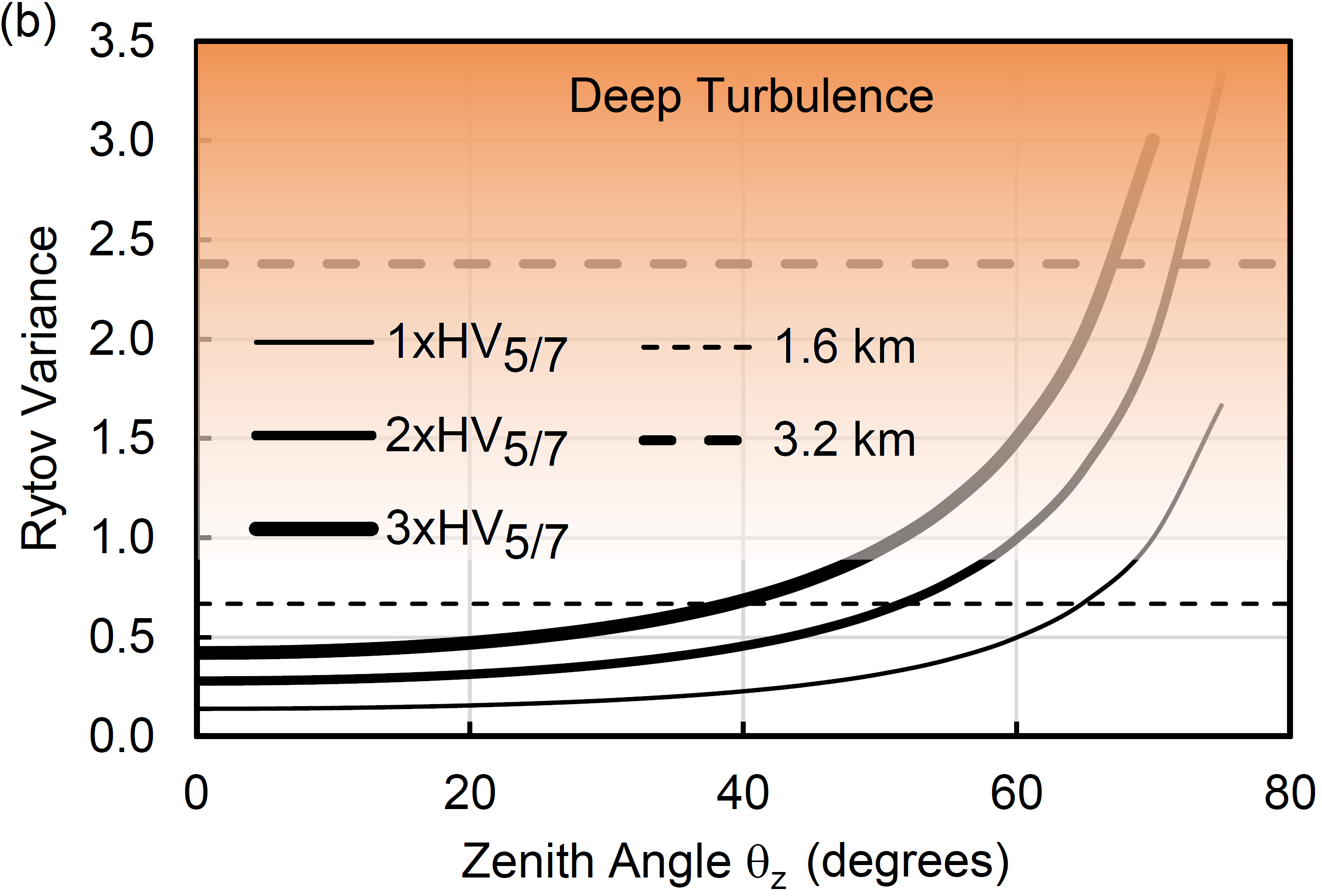}
	\caption{\label{fig:equiv_lengths}
Summary of theoretical calculations showing the 1.6-km propagation path produces spatial turbulence characteristics representative of slant-path propagation including (a) equivalent horizontal propagation distances required to produce comparable spatial coherence and scintillation to that introduced by slant-path propagation over a range of zenith angles and (b) Rytov variance vs. zenith angle for three different strengths of turbulence with red shaded region showing onset of deep turbulence effects.  Horizontal dashed lines indicate Rytov variance for 1.6-km and 3.2-km horizontal propagation distances. The 1.6-km horizontal path introduces scintillation effects commensurate with slant path propagation over a wide range of zenith angles but without entering the realm of deep turbulence typically not encountered in slant-path propagation.   
	}
\end{figure}

Figure ~\ref{fig:equiv_lengths}(a) shows the equivalent horizontal path distances calculated from Eqs.~\ref{eq:LR0} and \ref{eq:LRYTOV} and plotted for zenith angles ranging from $\theta_z = 0^{\circ}$ to 75$^{\circ}$.  
Introducing $r_0$ and Rytov variances comparable to those near zenith only require about 400 m and 700 m of horizontal propagation, respectively.  
Achieving values comparable to those at $\theta_z = 60^{\circ}$ requires about 800 m and 1.4 km, respectively.  
The horizontal dashed line identifies the 1.6-km propagation path length chosen for this field experiment.  
The model indicates that 1.6 km yields $r_0$ and Rytov variances corresponding to $\theta_z \approx 77^{\circ}$ and 65$^{\circ}$, respectively.  
Hence, a 1.6-km propagation path should introduce $r_0$ and Rytov variances that are at least as challenging as those encountered in slant-path turbulence over much of the sky hemisphere where satellite passes occur.

Over a sufficiently long propagation distance, turbulence-induced wavefront errors give rise to transverse amplitude variations.
In deep turbulence, these amplitude variations lead to intensity nulls that are not only less representative of slant-path propagation but also pose problems for the Shack-Hartmann wavefront sensor (SHWFS) that is commonly used in astronomical AO.  
Figure~\ref{fig:equiv_lengths}(b) shows the calculated Rytov variance for 1$\times$HV$_{5/7}$, 2$\times$HV$_{5/7}$, and 3$\times$HV$_{5/7}$ turbulence strengths over the range $0^{\circ} \leq \theta_z \leq 75^{\circ}$.  
Deep turbulence effects are known to appear when the Rytov variance reaches a value of about 0.8 and increase thereafter \cite{barchers2002evaluation}.  
Note that the Rytov variance, or log-intensity variance, is 4-times the often quoted Rytov number, or log-amplitude variance \cite{sasiela2012electromagnetic}.  
The shaded red region in Fig.~\ref{fig:equiv_lengths}(b) indicates the onset and growth of deep turbulence effects.  
The horizontal dashed lines indicate the calculated Rytov variance for 1.6-km and 3.2-km horizontal propagation distances.
The model indicates that the 1.6-km horizontal path introduces scintillation effects commensurate with slant path propagation over a wide range of zenith angles but without entering the realm of deep turbulence.   
Figure~\ref{fig:equiv_lengths}(b) illustrates the compatibility between slant-path turbulence, the 1.6-km propagation site, and the SHWFS approach chosen for this field experiment.  
Evaluation of scintillation data from both 1.6-km and 3.2-km propagation sites confirmed the shorter 1.6-km site was a more relevant representation of slant-path turbulence.  
This conclusion, based on both analysis and measurement, is consistent with 1.6-km path that was used to calibrate the OPALS AO system prior to the successful demonstration of optical communications from LEO \cite{wright2015adaptive}.   

\begin{figure}[t]
	\includegraphics[width=1\columnwidth]{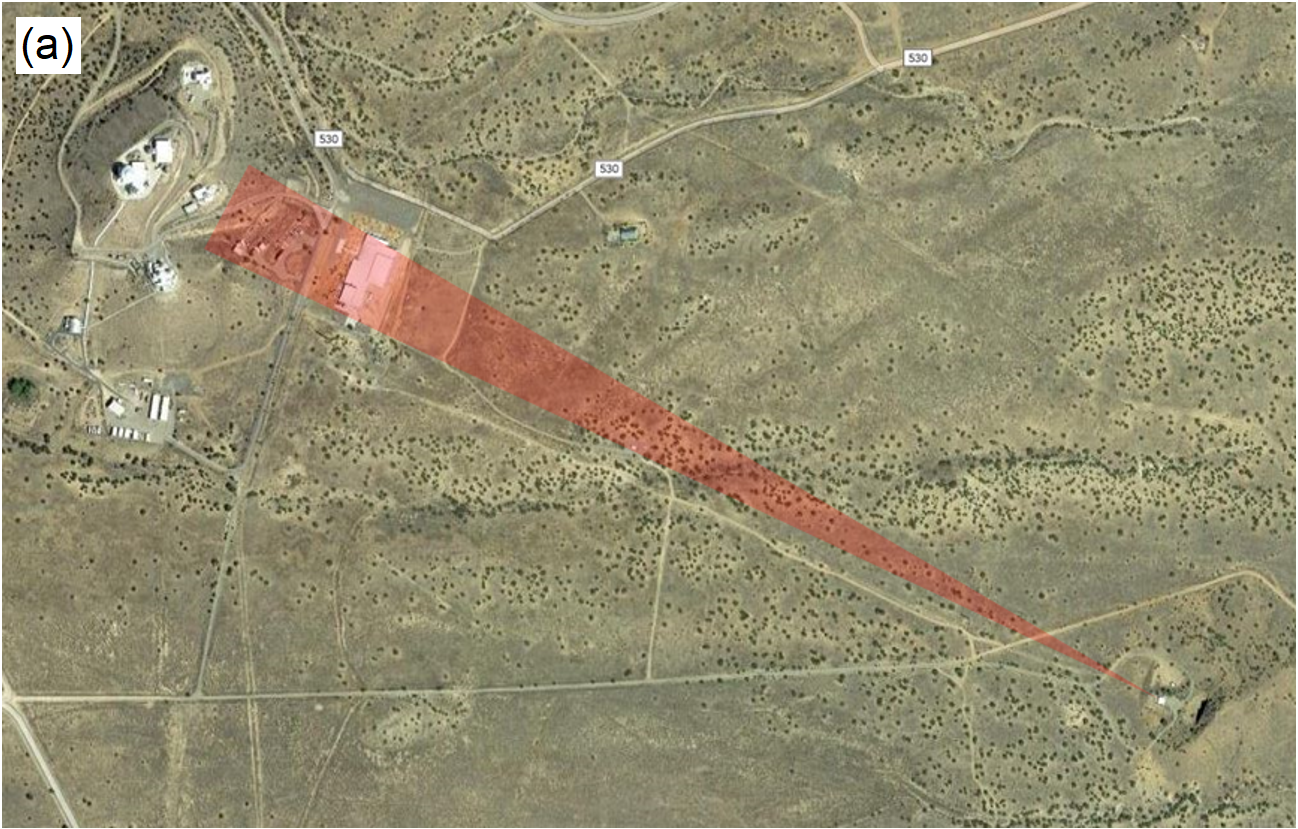}\\
      \vspace{0.25cm} 
	\includegraphics[width=1\columnwidth]{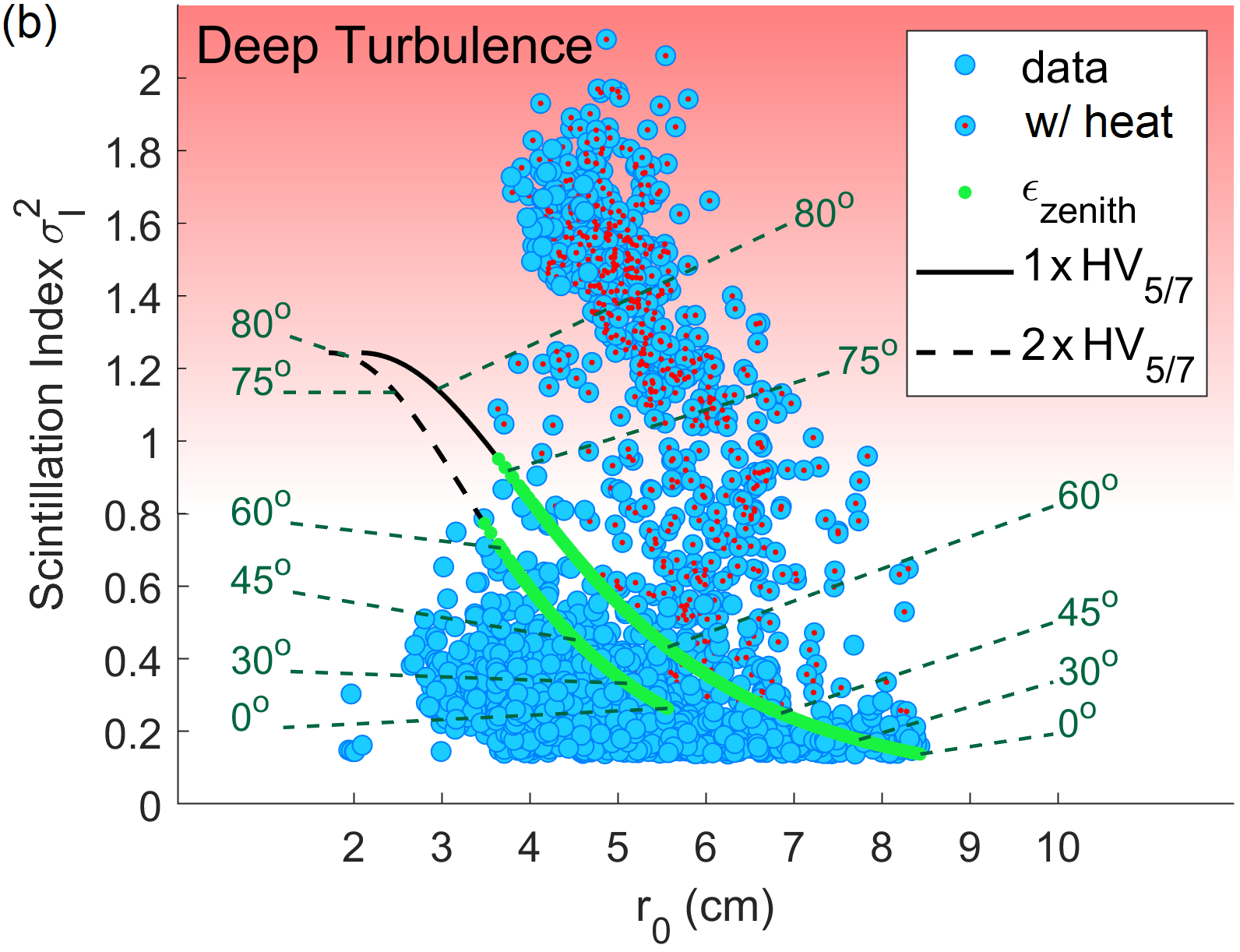}
	\caption{\label{fig:overhead_pairs}
Experimental confirmation that the 1.6-km propagation distance introduces turbulence characteristics relevant to slant-path propagation including, (a) an aerial photograph from Google Maps showing propagation from transmitter to receiver along northwest horizontal path and, (b) plot of the measured turbulence parameters $[ r_0, \sigma_I^2 ]$ in blue with corresponding equivalent zenith angles $\varepsilon_{\mathrm{zentih}}$ in green. Blue dots with red inset indicate data that was acquired with a heat source under the propagation path near the transmitter. The solid (dashed) line shows the theoretical 1$\times$HV$_{5/7}$ (2$\times$HV$_{5/7}$) turbulence profile representing a slant-path channel over the range of zenith angles.   
	}
\end{figure}

The field site is located at the Starfire Optical Range (SOR), Kirtland AFB, NM in the Southwestern United States.  
The site includes transmitter and receiver facilities located on hillsides that are separated by approximately 1.6 km. 
Figure~\ref{fig:overhead_pairs}(a) shows an aerial photograph from Google Maps illustrating the transmitter and receiver locations and the propagation path which is approximately 10 m above the desert floor.  
Atmospheric turbulence in the quantum channel was characterized from data acquired by the SHWFS that was integral to the channel and operated at 2-kHz.  
An independent measurement was provided by a Scintec BLS 900 scintillometer.  

Figure~\ref{fig:overhead_pairs}(b) compares measured $[ r_0, \sigma_I^2 ]$ pairs to those calculated for 1$\times$HV$_{5/7}$ and 2$\times$HV$_{5/7}$ turbulence strengths.  
The solid and dashed lines show the calculated pairs with coordinates corresponding to specific zenith angles labeled.  
Measured $[ r_0, \sigma_I^2 ]$ pairs are plotted for $\lambda=$ 780 nm as solid blue circles.  
The displacement of the two curves in Fig.~\ref{fig:equiv_lengths}(a), indicates that reproducing scintillation values for a given zenith angle requires a somewhat longer horizontal propagation path than that required to reproduce $r_0$.  
This was addressed by adding a heat source under the channel near the transmitter to increase scintillation in some of the data sets.  
Measured $[ r_0, \sigma_I^2 ]$ pairs acquired with the addition of a heat source under the beam path near the transmitter are designated by a red dot at the center of the blue circles.  
The measured $r_0$ and $\sigma_I^2$ span the full range of those calculated for 1$\times$HV$_{5/7}$ and 2$\times$HV$_{5/7}$ turbulence within $0^{\circ} \leq \theta_z \leq 89^{\circ}$ and even provide additional data in the realm of deep turbulence.  
Turbulence conditions vary considerably over the course of a day and include exceptionally mild turbulence during the quiescent periods near sunrise and sunset.  
Data points acquired with $\sigma_I^2 <$ 0.14 and $r_0 >$ 8.5 cm do not correspond to any downlink scenario in the 1$\times$HV$_{5/7}$ and 2$\times$HV$_{5/7}$ models and are omitted from both the plot and the data analysis that follows. 

Both turbulence and quantum data were processed in 10-s intervals, which were found to be optimum for specifying turbulence parameters while minimizing statistical variations in quantum detection events due to finite sample sizes.  
The $r_0$ values were derived from SHWFS measurements through the slope discrepancy independent of the open or closed state of the AO control loop \cite{brennan2003anatomy}.  
The measured $\sigma_I^2$ values are averages based on the variance of the intensity within each subaperture \cite{vedrenne2007c}.  
The models, which assume constant $C_n^2$ over the horizontal path, predict a maximum value for the saturated scintillation index of $\sigma_I^2 \approx 1.7 $ (see appendix~\ref{sec:appendixA} for relevant equations). 
Experimentally, somewhat higher values are recorded and the measurements of $r_0$ and $\sigma_I^2$ are not correlated as strongly as the model would indicate.  
These experimental variations occur as a consequence of variations in $C_n^2$ across the propagation path.  

Each data point can be projected onto the HV$_{5/7}$ curves and assigned a maximum zenith angle $\varepsilon_{\mathrm{zentih}}$ within which the field-site conditions were more demanding than those in the HV$_{5/7}$ models.  
More specifically, we define $\varepsilon_{\mathrm{zentih}}$ to be the largest angle for a given measured $[ r_0, \sigma_I^2 ]$ pair where $ r_0 \leq r_0$(HV$_{5/7})$ and $\sigma_I^2 \geq \sigma_I^2$(HV$_{5/7}$).  
Practically, one can project points above the curve down and points to the left of the curve to the right to establish $\varepsilon_{\mathrm{zentih}}$.  
The green points on the HV$_{5/7}$ curves represent all of the identified $\varepsilon_{\mathrm{zentih}}$ values from the processed data and for 1$\times$HV$_{5/7}$ and 2$\times$HV$_{5/7}$ turbulence span the ranges $0^{\circ}  \leq \varepsilon_{\mathrm{zentih}} \leq 76^{\circ}$ and $0^{\circ} \leq \varepsilon_{\mathrm{zentih}} \leq 63^{\circ}$, respectively.

\subsection{Simulating daytime sky noise in a terrestrial path}

In daytime, background photons due to scattered sunlight are the dominant source of channel noise and QBER.  
We introduced a range of background radiance that is comparable to that experienced by a ground telescope tracking a satellite across the daytime sky.  
At 780-nm wavelength, daytime sky radiances over most of the sky hemisphere lie within the range $2<H_b<100$ $\mathrm{W} / ( \mathrm{m}^2 \, \mathrm{sr} \, \mu \mathrm{m})$ \cite{gruneisen2014adaptive, nordholt2002present, gruneisen2015modeling}. 
The natural channel radiances occurring with horizontal propagation at the field site only partially overlap with this range.  
This was addressed by adding an unpolarized white-light source beside the transmitter and outside the receiver FOV to introduce additional background light into the channel via scattering.  

 \begin{figure*}[t!]
	\includegraphics[width=.82\textwidth]{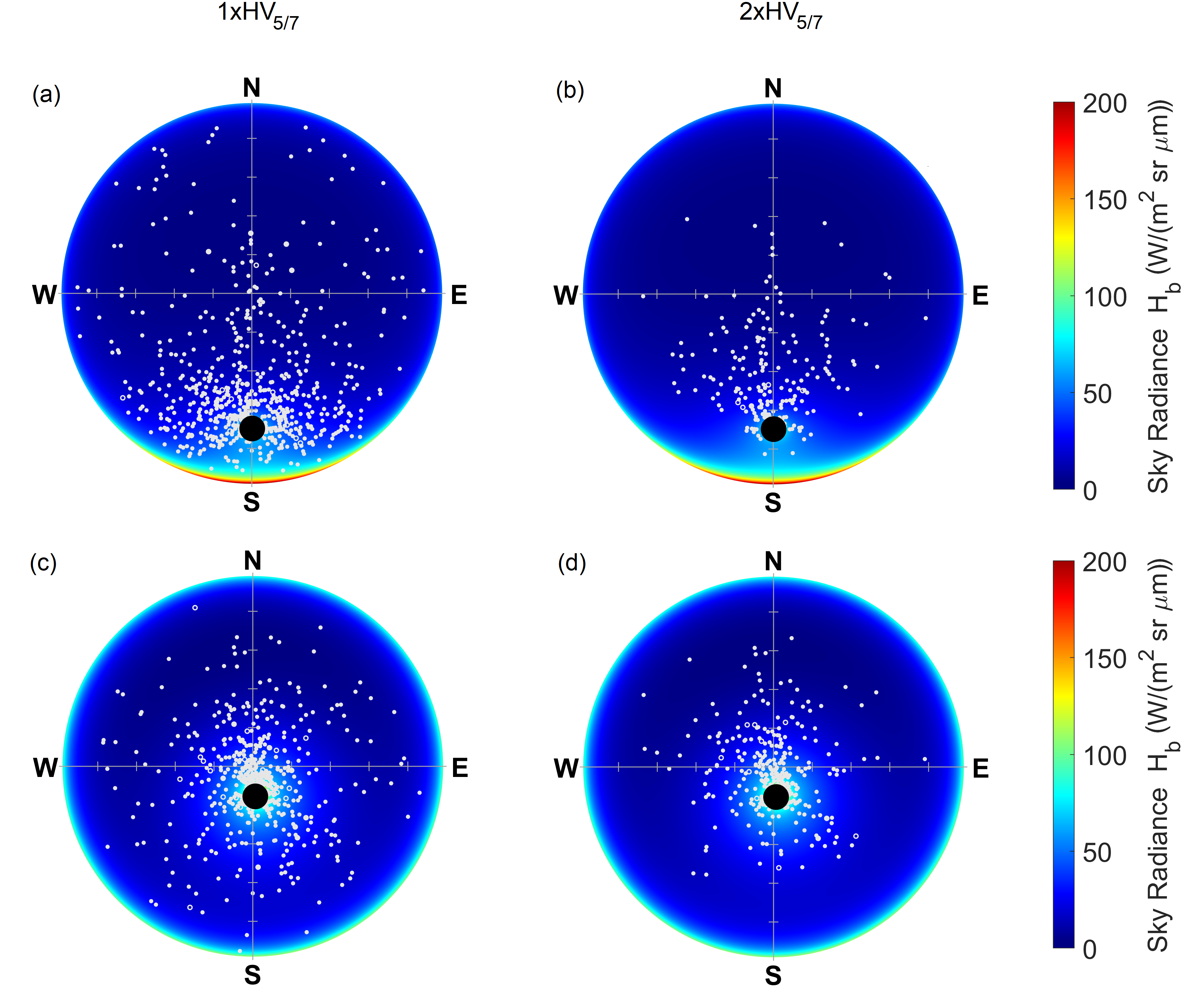}\\
	\caption{\label{fig:hemi_plots}
Hemispherical plots showing sky angles for which daytime atmospheric channels were simulated in the field experiment.  Color scale shows sky radiances for the winter solstice (top) and summer solstice (bottom) with the sun position obscured by a black circle subtending 14$^{\circ}$.  Experimental data sets with comparable atmospheric scintillation, spatial coherence, and background radiance are shown as white circles for both open-loop (open circles) and closed-loop (solid circles) cases for both 1$\times$HV$_{5/7}$ (left) and 2$\times$HV$_{5/7}$ (right) turbulence strengths.   
	}
\end{figure*}
Figure~\ref{fig:hemi_plots} illustrates the sky angles for which conditions in the field experiment approximated both daytime sky radiance and slant-path turbulence conditions.  
The large circles are hemispherical plots in which the background color map gives the noon-time sky radiance predicted by radiative transfer modeling and shared from Ref.~\cite{gruneisen2015modeling}.  
The region about the sun angle is represented by a black circle subtending 14$^{\circ}$.  
The top row illustrates the winter solstice where the peak sun angle is only 31$^{\circ}$ above the horizon.  
In this case, the atmospheric path near the sun angle is relatively long leading to both elevated sky brightness and increased turbulence.  
The bottom row illustrates the case of the summer solstice where the sun achieves a much higher angle of about 78$^{\circ}$ above the horizon.  
In this example, the sun angle is within 60$^{\circ}$ of zenith where turbulence conditions are the most benign.  
Results obtained with $\varepsilon_{\mathrm{zentih}}$ falling on the 1$\times$HV$_{5/7}$ turbulence profile in Fig.~\ref{fig:overhead_pairs}(b) are shown on the left.  
Those obtained with $\varepsilon_{\mathrm{zentih}}$ on the 2$\times$HV$_{5/7}$ profile are shown on the right.  
The white circles represent experimental data points showing combinations of $\varepsilon_{\mathrm{zentih}}$ and $H_b$ under which data were recorded.  
More specifically, for a given data point, $\varepsilon_{\mathrm{zentih}}$ defines a circle of constant $[ r_0, \sigma_I^2 ]$ inside of which the point is placed such that the measured $H_b$ matches the calculated sky radiance.  
As the sky radiance is not symmetric, most data points find a single match on the hemispherical plots within a degree of their $\varepsilon_{\mathrm{zentih}}$.  

Solid white circles represent data sets acquired with the AO control loop closed and open circles represent open-loop data sets.
In the open-loop state, the DM was optically flat but the tip/tilt loop remain closed. 
Figure~\ref{fig:hemi_plots} illustrates the sky angles for which our scintillation, spatial coherence, and sky radiance conditions at the SOR 1-mile range were at least as demanding as those predicted for 1$\times$HV$_{5/7}$ and 2$\times$HV$_{5/7}$ turbulence profiles over the daytime sky hemisphere.  
The clustering of data points near the black circle indicate that the very challenging region of high background near the sun angle was substantially explored.

\subsection{Temporal characteristics of turbulence}

The temporal rate of change in the wavefront error is quantified through the Greenwood frequency $f_\mathrm{G}$ \cite{greenwood1977bandwidth}. 
Compensation of dynamic wavefront errors is most effective when the closed-loop control bandwidth of the AO system $f_c$ exceeds $f_\mathrm{G}$.  
The 130-Hz closed-loop bandwidth of the AO system in the field experiment was more than twice the maximum $f_\mathrm{G}$ of 62 Hz that was measured in the quantum channel.  
In the analysis that follows, we place these field-experiment parameters in the context of actual slant-path turbulence dynamics.

\begin{figure}[t!]
	\includegraphics[width=.98\columnwidth]{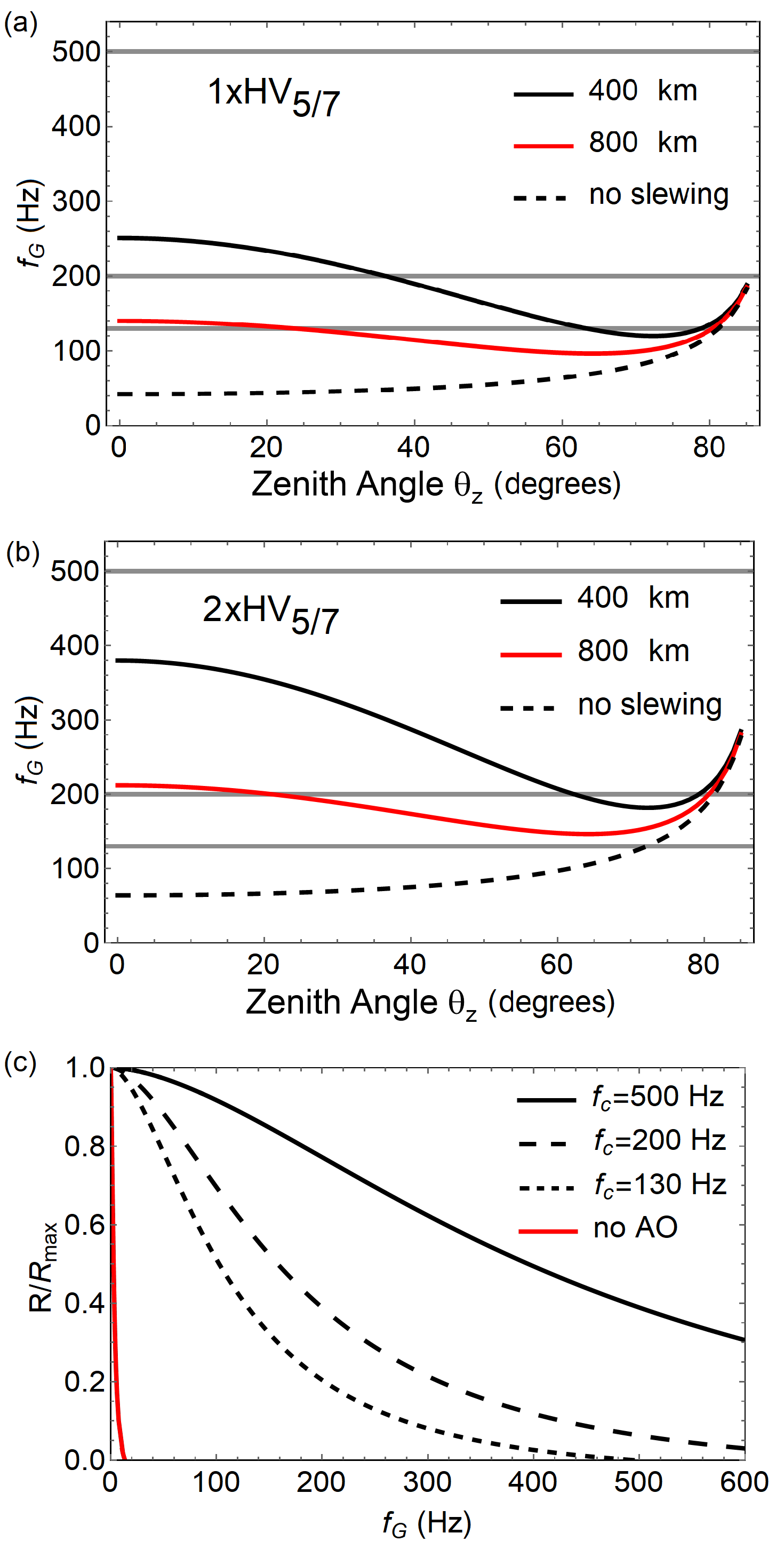}
	\caption{\label{fig:relevancy}
Theoretical plots demonstrating the relevancy of the field-experiment AO system to actual slant-path turbulence compensation and satellite engagements. In (a) and (b) the Greenwood frequency $f_{\mathrm{G}}$ is calculated with and without slewing and plotted vs. zenith angle for two turbulence strengths. 
In (c), the normalized decoy-state QKD bit-yield is plotted as a function of $f_\mathrm{G}$ without the benefit of AO (red line) and for three AO control-loop bandwidths (black, dashed, and solid lines). This shows that any of these bandwidths could be enabling for satellite-to-Earth downlinks.  
	}
\end{figure}

Temporal fluctuations intrinsic to an atmospheric channel occur due to wind.  
When a ground telescope tracks a moving satellite through turbulence, slewing leads to additional dynamics in the turbulence-induced wavefront error that can increase $f_\mathrm{G}$.  
Previously, we presented analyses demonstrating the efficacy of 200-Hz and 500-Hz bandwidth AO systems for satellite-Earth QKD from LEO including scenarios where $f_c < f_\mathrm{G}$ \cite{gruneisen2016adaptive, gruneisen2017modeling}.  
These analyses demonstrate that $f_c$ does not denote a sharp cutoff frequency with respect to $f_\mathrm{G}$.  
A useful degree of AO correction can be achieved with $f_c < f_\mathrm{G}$ and degradation in AO compensation is typically graceful as $f_\mathrm{G}$ approaches and exceeds $f_c$.  
Appendix~\ref{sec:appendixB} reviews equations for calculating $f_\mathrm{G}$ for circular orbits with overhead passes that intersect zenith.  
Figure~\ref{fig:relevancy}(a) shows $f_\mathrm{G}$ calculated for 1$\times$HV$_{5/7}$ turbulence over a range of zenith angles for 400- and 800-km circular orbits.  
The dashed line is the $f_\mathrm{G}$ that is intrinsic to the atmospheric channel calculated without the effects of slewing.
The horizontal gray lines indicate the 130-, 200-, and 500-Hz AO systems considered in this field experiment and these earlier analyses.  
Figure~\ref{fig:relevancy}(b) shows the corresponding results for 2$\times$HV$_{5/7}$ turbulence.

The 62-Hz $f_\mathrm{G}$ measured in the field experiment is comparable to or exceeds the intrinsic slant-path $f_\mathrm{G}$ for $1\times$HV$_{5/7}$ turbulence within a 116$^{\circ}$ cone about zenith.   
The 130-Hz control-loop bandwidth of the AO system exceeds the intrinsic $f_\mathrm{G}$ for both 1$\times$HV$_{5/7}$ and 2$\times$HV$_{5/7}$ turbulence within a 140$^{\circ}$ cone about zenith.  
With slewing at a rate to track a satellite in an 800-km orbit through 1$\times$HV$_{5/7}$ turbulence, the 130-Hz bandwidth is comparable to or greater than all $f_\mathrm{G}$ values within a 160$^{\circ}$ cone angle.  
With either stronger 2$\times$HV$_{5/7}$ turbulence or the lower 400-km orbit, $f_\mathrm{G}$ increases and exceeds the 130-Hz AO bandwidth.  
The higher 200-Hz bandwidth is more compatible with both orbit altitudes in 1$\times$HV$_{5/7}$ turbulence and the 500-Hz control-loop bandwidth exceeds $f_\mathrm{G}$ for all cases considered.  

The impact of AO bandwidth on satellite-Earth QComm can be quantified by calculating a QKD bit-yield probability following the approach presented in \cite{lanning2021quantum}.  
More specifically, the bit yield $R$ is calculated for the vacuum-plus-decoy-state QKD protocol \cite{ma2005practical} as a function of the channel efficiency which is heavily influenced by the FS transmission.  
For cases with AO compensation, the FS transmission is a function of the effective, or residual, $r_0$ which is a function of $f_c$ and $f_G$.  
Equations for the decoy-state QKD bit yield are reviewed in Appendix~\ref{sec:appendixC} and the methodologies for calculating the residual $r_0$ and FS transmission can be found in \cite{lanning2021quantum}.  
In this example, we assume source characteristics given by $\lambda=780$ nm, a signal mean photon number (MPN) $\mu=0.6$, and a decoy-state MPN $\nu=0.1$.  
The link geometry is defined by $D_{\mathrm{T}}=20$ cm, $D_{\mathrm{R}}=1$ m, and $z=700$ km.   
Contributions to the channel efficiency include those due to the aperture coupling efficiency associated with Gaussian beam propagation, $\eta_{\mathrm{geo}}=0.078$ (or $-$11 dB), atmospheric scattering and absorption, $\eta_{\mathrm{trans}}=0.9$, receiver telescope optics, $\eta_{\mathrm{rec}}= 0.5$, spectral filtering, $\eta_{\mathrm{spec}}=0.9$, and each of the four APD detectors, $\eta_{\mathrm{det}} = 0.6$.  
The turbulence-related losses at the field stop $\eta_{\mathrm{FS}}$ are calculated for each combination of $f_c$ and $f_G$.  
Contributions to background include the sky radiance, $H_b=25$ $\mathrm{W} / ( \mathrm{m}^2 \, \mathrm{sr} \, \mu \mathrm{m})$, the detector dark count rate, $f_{\mathrm{dark}}=250$ Hz, the background rate, $e_0=0.5$, and polarization crosstalk, $e_d=0.01$.  
Background filtering is introduced through temporal filtering, $\Delta t=1$-ns, spectral filtering, $\Delta \lambda=1$-nm, and the D-L FOV, $\Omega_{\mathrm{FOV}}$.  
The efficiency of error correction is taken to be a constant, $f(E_{\mu}) =1.22$.

Figure~\ref{fig:relevancy}(c) shows the calculated bit-yield probability as a function of $f_\mathrm{G}$ for the case without AO and for the cases with 130-, 200-, and 500-Hz AO control-loop bandwidths.  
The case without AO is shown in red.  
For the system parameters considered in this paper, the calculation shows AO is necessary to achieve a viable quantum channel over any range of realistic $f_\mathrm{G}$.  
These results also show that any of the three bandwidths, including the 130-Hz system demonstrated in this field experiment, result in a viable quantum channel over a useful range of $f_\mathrm{G}$.  
In all cases, the bit yield declines with increasing $f_\mathrm{G}$.

The plots in Fig.~\ref{fig:relevancy} are presented for context.  
The slew dynamics that affect AO bandwidth requirements can be considered separately from the intrinsic dynamics of the atmosphere.  
These effects have been addressed previously by others and high-bandwidth AO systems for the most challenging LEO applications have already been demonstrated. 
For this field experiment performed with a stationary transmitter, it was not necessary develop a higher bandwidth system.

\subsection{Receiver aperture size and aperture-to-aperture coupling efficiency}

At the receiver station, light was collected by a commercial 35-cm Schmidt-Cassegrain telescope such that $D_{\mathrm{R}} > r_0$ for all $r_0$ within the range 2 cm to 8.5 cm as are shown for recorded data sets in Fig.~\ref{fig:overhead_pairs}(b).  
Under this condition, higher-order wavefront errors are a factor affecting FS transmission efficiency and the benefit of higher-order AO can be tested and demonstrated.  
The transmitter divergence was adjusted via defocusing to introduce approximately 1 mrad of full-angle divergence and 11 dB of aperture-to-aperture coupling loss.  
Additional divergence due to atmospheric turbulence over the 1.6-km path was estimated to be only about 0.2 dB.  
For the scenario with $D_{\mathrm{T}}$=20 cm, $D_{\mathrm{R}}$=1 m, $\lambda=780$ nm and local aerosol conditions, this represents an approximate range of orbit-altitude and zenith-angle pairs $[a,\theta_z]$ between $[400\,\mbox{km}, 60^{\circ}]$ and $[700\,\mbox{km}, 0^{\circ}]$.

\section{Integration of adaptive optics with a quantum communication system}\label{sec:section4}

The quantum communication system consisted of a transmitter (Alice) and receiver (Bob) that prepared, transmitted, and measured polarization-based weak-coherent-pulse qubits at 780-nm wavelength in rectilinear and diagonal bases of polarization.  
Alice also launched an 808-nm cooperative laser beacon to probe atmospheric turbulence and an 808-nm timing pulse that preceded each qubit by about 40 ns to facilitate temporal filtering of optical noise outside a $\Delta t = 1$-ns temporal window.  
Bob applied AO wavefront correction to qubit, beacon, and timing pulses prior to demultiplexing them. 

\begin{figure}[b]
	\includegraphics[width=1\columnwidth]{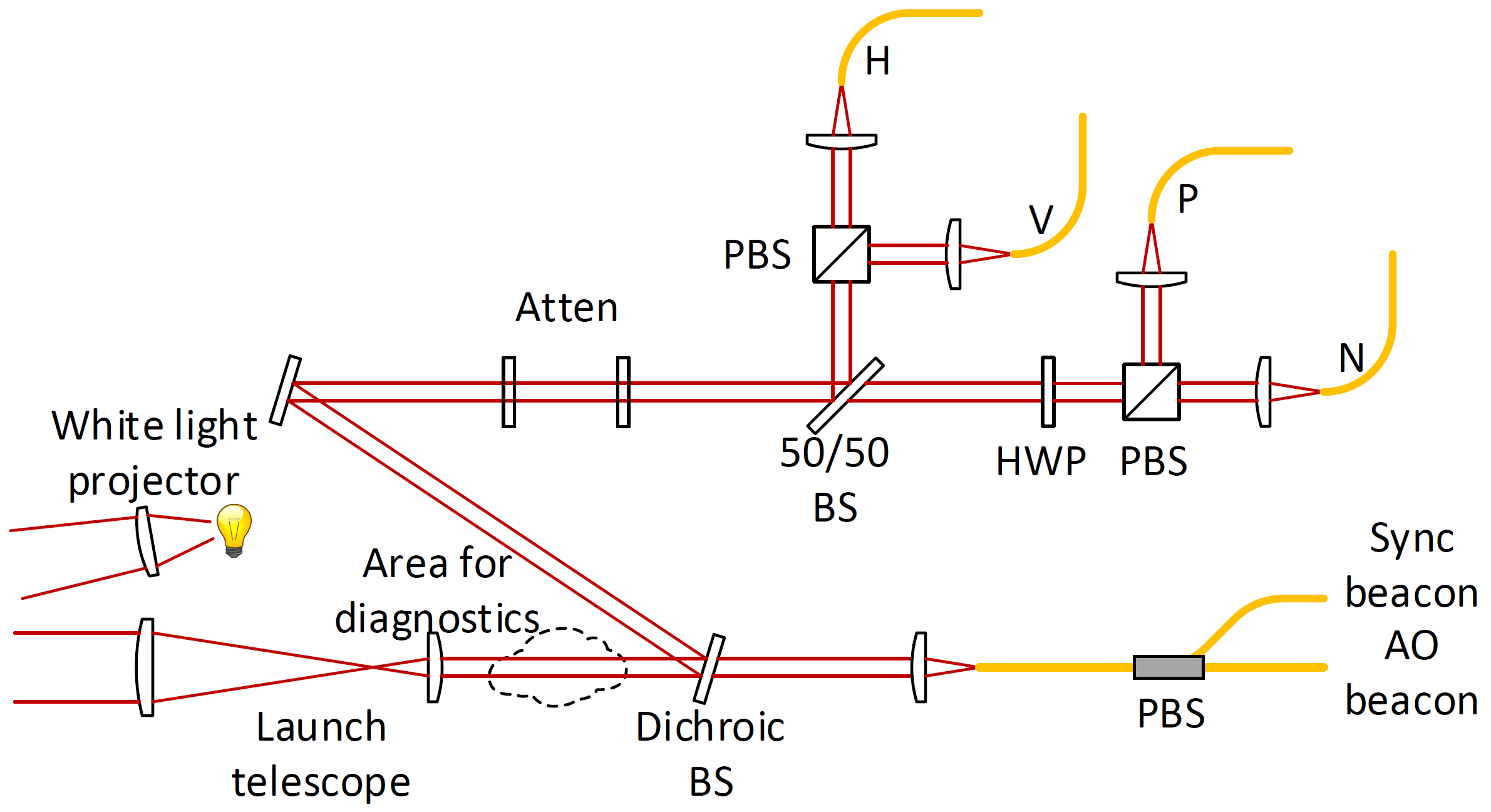}
	\caption{\label{fig:transmitter_hardware}
Partial schematic of quantum transmitter showing polarizing beam splitters (PBS), half-wave plate (HWP), and 50/50 beam splitter (BS) used to combine horizontal (H), vertical (V), positive 45$^{\circ}$ (P), and negative 45$^{\circ}$ (N) polarized signal pulses.  Neutral density filters (Atten) reduce mean photon numbers to less than unity.  A vertically polarized adaptive-optics (AO) beacon and horizontally polarized sync pulse are combined with a fiber-coupled PBS and combined with signal pulses at a dichroic BS.  A white-light projector, located outside the receiver FOV, introduces incoherent background light into the channel via atmospheric scattering.   
	}
\end{figure}

Alice transmitted qubits in 100-s sessions, each second comprised of a 1-MHz burst of 120,000 pulses of which 12,000 were vacuum-decoy pulses during which background was counted.  
The polarizations were equally distributed among the four rectilinear and diagonal polarizations.  
A partial illustration of the transmitter optical components is shown in Fig.~\ref{fig:transmitter_hardware}.  
The Alice computer controls four fiber-based Mach-Zehnder (MZ) modulators (not shown) in on/off configuration to generate 1.0-ns FWHM optical pulses with horizontal (H), vertical (V), positive 45$^{\circ}$ (P) and negative 45$^{\circ}$ (N) polarizations. 
These modulators are fed from a 4x fiber splitter illuminated by a single, 780-nm cw laser diode source.  
A fifth MZ modulator (not shown) is used as a fast amplitude adjuster to create either a signal MPN of $\mu \approx 0.6$ or a decoy-state MPN of $\nu \approx 0.1$.  
Vacuum-decoy pulses were achieved with all MZ modulators in the off configuration.  
The fiber outputs are collimated and combined in air with polarizing beam-splitter (PBS) cubes and a non-polarizing 50/50 beam-splitter (BS).  
A series of neutral density filters (Atten) attenuates the laser pulses to the desired MPN.  
The horizontally polarized 808-nm cw AO laser beacon is polarization multiplexed with the vertically polarized 808-nm 20-ns duration square heralding pulse and then combined with the 780-nm qubits with a dichroic BS.  
The launch portion of the transmitter is a simple Keplerian telescope built with commercial achromat lenses.  
The full-angle divergence at the 5-cm exit lens was 1 mrad and the beam radius was 9.5 mm.

A randomly-polarized white-light projector was added to the transmitter station to increase the channel radiance beyond the ambient level and was aligned across the line-of-sight from the launch telescope to the receiver telescope.  
This source was located outside the field of view of Bob and Rayleigh scattering along the path added randomly-polarized background light to the field of view of the photon counting detectors.  
The level of equivalent sky background was easily adjusted.  

\begin{figure}[t]
	\includegraphics[width=1\columnwidth]{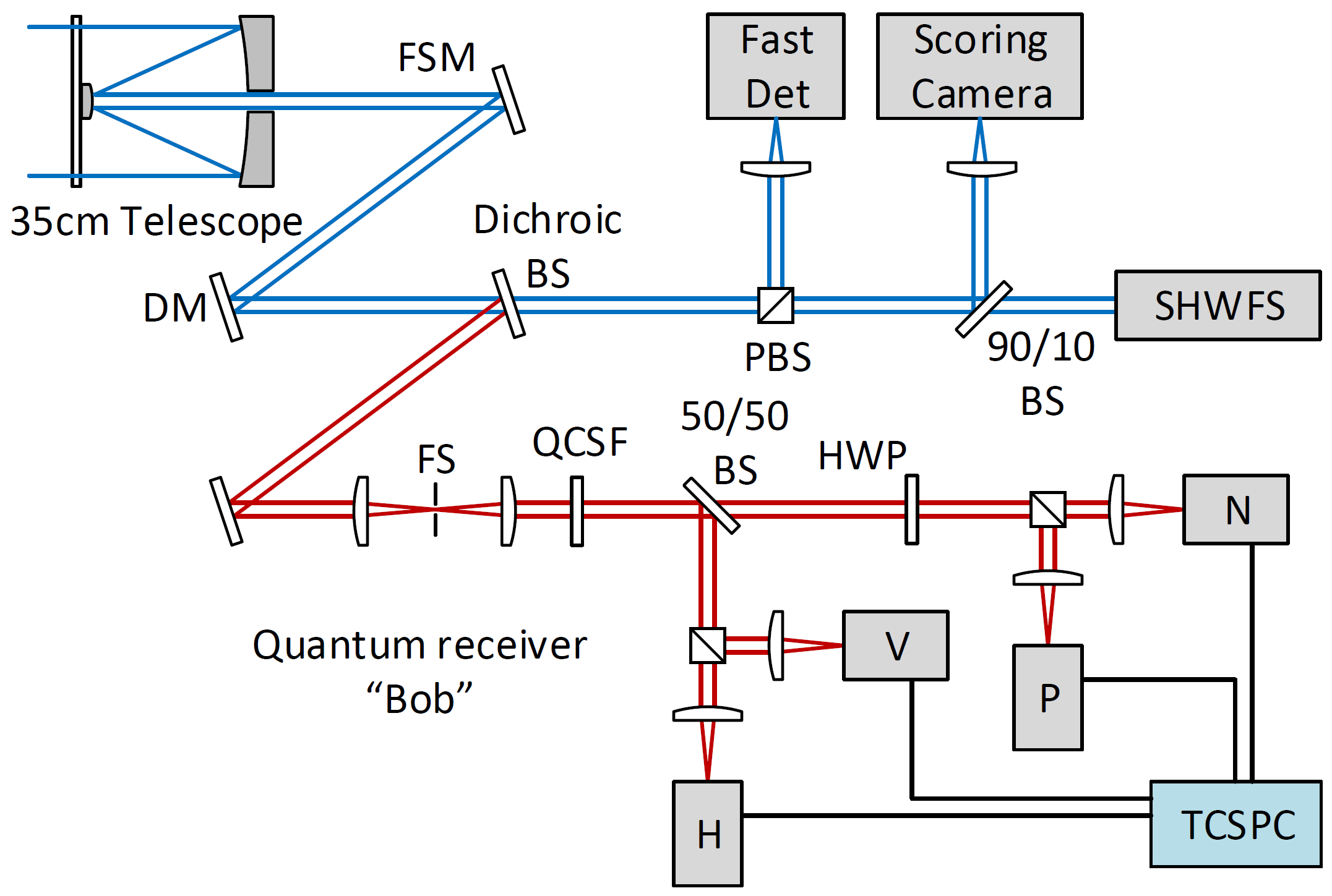}
	\caption{\label{fig:receiver_hardware}
Schematic of the quantum receiver including a fast-steering mirror FSM and deformable mirror DM.  A dichroic beam splitter BS separates the quantum signal from the AO beacon and sync pulse.  A polarizing beam splitter PBS, directs the sync pulse to a fast timing detector and a 90/10 BS directs beacon light to a Shack-Hartmann wavefront sensor SHWFS and scoring camera.  The quantum signal is filtered through a 30-$\mu$m diameter FS and a 1-nm bandwidth quantum-channel spectral filter QCSF before propagating to the standard arrangement for measuring polarization states in rectilinear and diagonal bases including a half-wave plate HWP for 45$^{\circ}$ polarization rotation. Detection events are registered by a time-correlated single photon event counter TCSPC.  
	}
\end{figure}
At the receiver station, light was collected by a commercial 35-cm Schmidt-Cassegrain telescope.  
Figure~\ref{fig:receiver_hardware} shows how captured light propagates to a FSM for atmospheric tip/tilt correction and a micro-electro-mechanical-system- (MEMS-) based DM for higher-order wavefront correction.  
A dichroic BS reflects the 780-nm qubit stream and transmits the 808-nm AO beacon and timing pulse which are subsequently de-multiplexed by a PBS.  
The heralding pulse is directed to a fast detector that generates the sync pulse for Bob's event counter. 
A 90/10 BS directs the AO beacon light to the SHWFS and an imaging “scoring” camera.  
For compatibility with the 35-cm aperture receiver telescope used in the field experiment, an 11$\times$11 element lenslet array was chosen for the SHWFS.  
This corresponds to a subaperture size of 3.2 cm at the entrance pupil of the telescope. 
The SHWFS camera ran at 2.0 kHz while maintaining excellent S/N in the sub-apertures.  

Referring to Fig.~\ref{fig:receiver_hardware}, one will see the qubit stream is focused through a 30-$\mu$m diameter circular FS, which serves as the spatial filter for optical noise, and passes through the 1.0-nm FWHM bandpass QCSF.  
The 30-$\mu$m FS restricts the quantum channel FOV to the 5.5 $\mu$rad D-L FOV and thereby permitted the use of the relatively broad 1.0-nm spectral filter.  
The remaining portion of Bob consists of the standard arrangement for measuring polarized photons in rectilinear and diagonal bases of polarization consistent with known prepare-and-measure protocols \cite{BB84QKD, ma2005practical}.  
A 50/50 non-polarizing BS randomly directs photons to either the rectilinear or diagonal measurement bases.  
In each basis, PBS cubes separate orthogonal polarizations and direct them to commercially-available Geiger-mode avalanche photodiodes (APDs).  
The four output signals are registered by a time-correlated single photon event counter (TCSPC) with picosecond resolution.

\begin{figure}[t]
	\includegraphics[width=1\columnwidth]{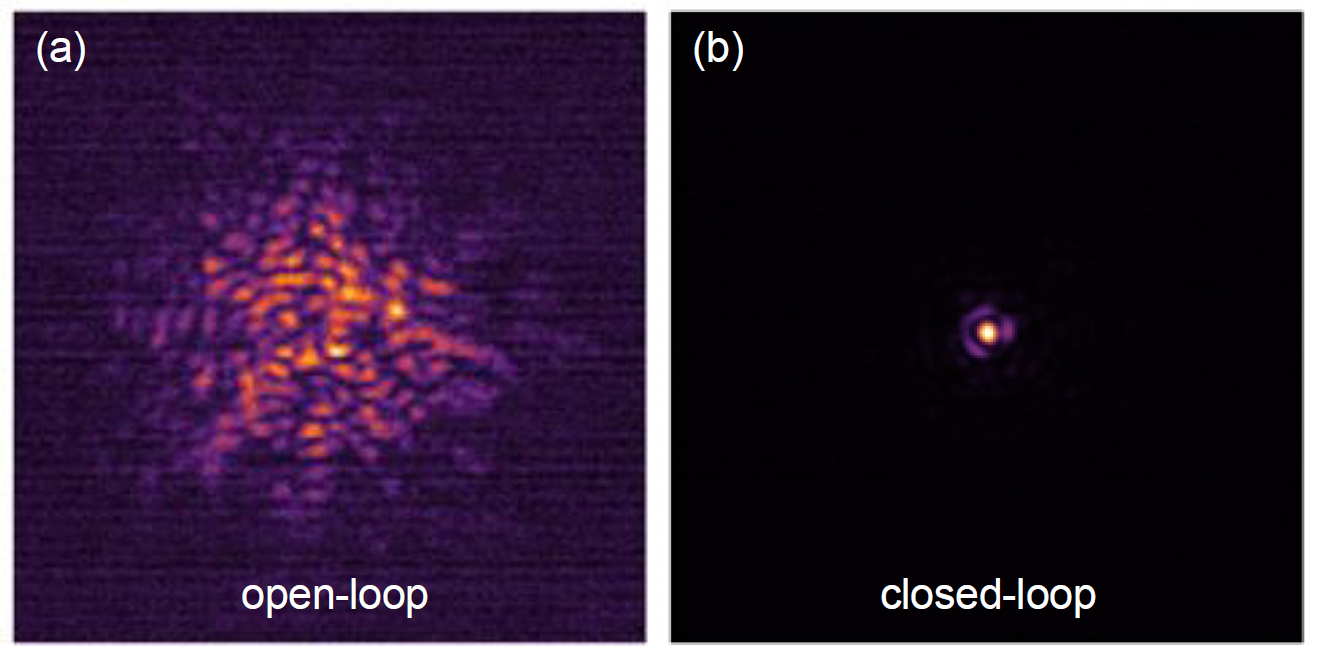}
	\caption{\label{fig:open_and_closed}
Auto-scaled scoring camera images demonstrating benefit  of AO to spatial filter size showing focal-plane spot sizes in (a) open-loop AO configuration and (b) closed-loop AO configuration resulting in 15-fold increase in peak intensity.   
	}
\end{figure}
The optical combination of the 35-cm diameter receiver aperture, the 30-$\mu$m diameter FS, and the spectral filter formed an extended radiometer for measuring background radiance when transmitting vacuum-decoy pulses.  
The integration time per pulse was increased from the 1-ns integration time for non-vacuum pulses to 78 ns in order to increase the sensitivity of background measurements.  
The number of background photons $N_b$ recorded on the four Bob detectors were converted into an equivalent sky radiance $H_b$ in $\mathrm{W} / ( \mathrm{m}^2 \, \mathrm{sr} \, \mu \mathrm{m})$ by solving Eq.~\ref{eq:NUMBERDL} for $H_b$ with $\Delta t$ taken to be the total time of background counting within the 10-s interval in which the atmospheric and quantum data were processed.   

The low-latency 12$\times$12 element MEMS DM was driven by the filtered, reconstructed wavefront and similarly, the 5-cm diameter FSM was driven by the full-pupil tilts.  
\begin{figure*}[t!]
	\includegraphics[width=0.98\textwidth]{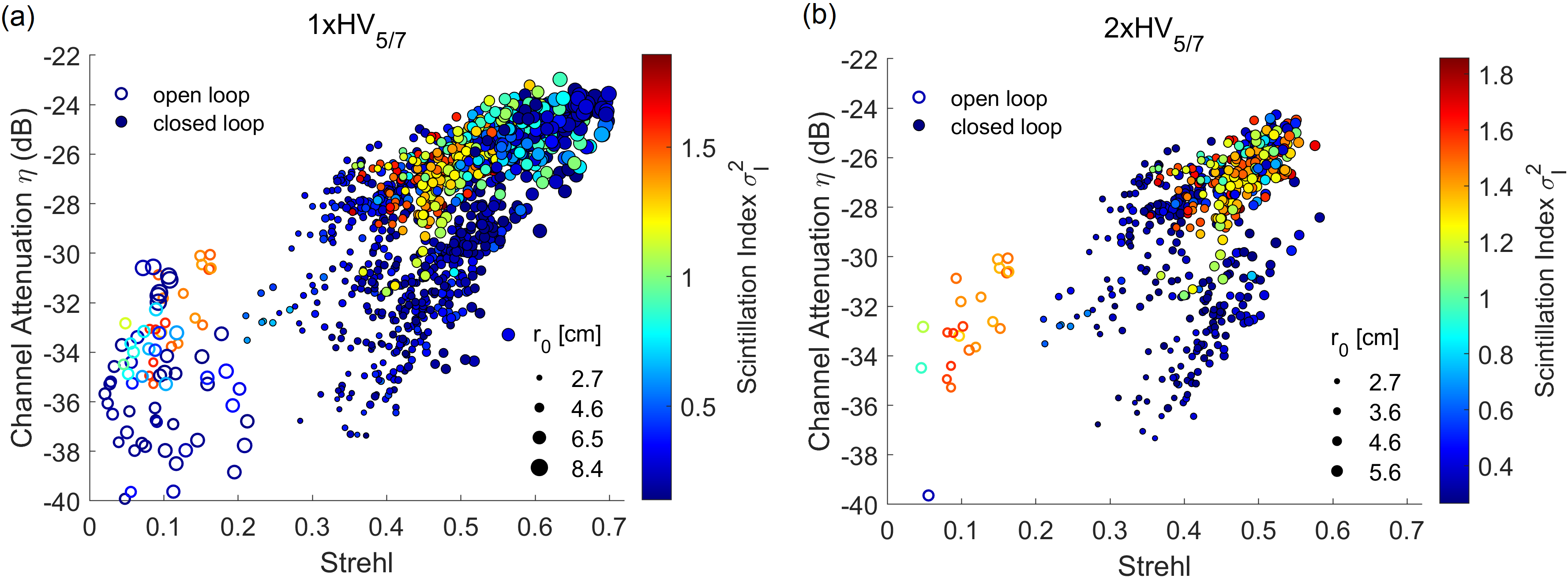}
\caption{\label{fig:eta}
Experimental results showing AO can increase quantum channel efficiencies by more than 10 dB while spatially filtering sky noise at the diffraction limit.  Channel attenuation is plotted in dB versus system Strehl for (a) 1$\times$HV$_{5/7}$ turbulence data sets and (b) 2$\times$HV$_{5/7}$ data sets for open-loop (open circles) and closed-loop (solid circles) cases with $r_0$ and $\sigma_I^2$ represented by circle size and color scale, respectively. 
	}
\end{figure*}
The 3-dB error rejection bandwidth of the AO loop was 130 Hz which was greater than the range of Greenwood frequencies encountered during data collection.  
Notably, all of the real-time AO loop and FSM control was accomplished with a single PC running tailored AO software.  
The computer was also networked for data storage: data streams associated with the SHWFS and AO loop, including WFS camera frames, gradients, reconstructed phase, servo and DM commands, were saved for post processing. 
The scoring camera was optically conjugate to the FS allowing a means of quantifying the quantum channel FS transmission efficiency and AO performance under all atmospheric conditions.  
Representative scoring camera frames for open- and closed-loop configurations are shown in Figs.~\ref{fig:open_and_closed}(a) and \ref{fig:open_and_closed}(b), respectively.  
The FS was sized to transmit only the central lobe in Fig.~\ref{fig:open_and_closed}(b).

\section{Experimental Results}\label{sec:section5}

The quantum channel efficiency $\eta$ was characterized through measurements of quantum parameters performed with APDs in the quantum channel.  
Measured parameters included the signal MPN $\mu$, measured at the output of the transmitter, and the signal-state gain $Q_\mu$ and the background probability $Y_0$, measured in the quantum receiver.  
The quantum channel efficiency was calculated from the relationship \cite{ma2005practical},
 \begin{equation}\label{eq:QMU}
Q_\mu = Y_0 + 1 - e^{-\eta \mu}.
\end{equation}
Contributions to $\eta$ that were characterized separately include the efficiencies associated with aperture coupling, measured to be $\eta_{\mathrm{geo}} \approx $ 0.08, atmospheric scattering and absorption, $\eta_{\mathrm{trans}} \approx $1.0, receiver telescope optics, $\eta_{\mathrm{rec}} = $ 0.49, spectral filtering, $\eta_{\mathrm{spec}} = $ 0.92, and each of the four APD detectors, $\eta_{\mathrm{det}} \approx $ 0.6.  
The turbulence-related losses at the FS $\eta_{\mathrm{FS}}$ varied with atmospheric realizations from a low of about 2$\%$ to as much as 63$\%$.  
In the most demanding cases, under-resolved wavefronts and scintillation reduce the accuracy 
of SHWFS measurements and, correspondingly, the fidelity of wavefront compensation \cite{roggemann2018imaging}.  

AO performance was quantified through classical measurements of Strehl given in the Marechal approximation by $S = \exp (- \sigma^2)$ where $\sigma^2$ is the phase variance calculated from SHWFS data \cite{sasiela2012electromagnetic}. 
Figure~\ref{fig:eta} shows the measured $\eta$ in dB loss versus Strehl with $r_0$ represented by circle sizes and the scintillation index $\sigma_I^2$ given by the color scale.  
Open- and closed-loop data are represented by open circles and solid circles, respectively.  
Results obtained with $\varepsilon_{\mathrm{zenith}}$ falling on the 1$\times$HV$_{5/7}$ turbulence profile in Fig.~\ref{fig:overhead_pairs}(b) are shown on the left.  
Those obtained with $\varepsilon_{\mathrm{zenith}}$ on the 2$\times$HV$_{5/7}$ profile are shown on the right.  
Overall, closing the AO loop improved the range of system Strehl from $0.024 \leq S \leq 0.21$ to $0.21 \leq S \leq 0.70$ with the maximum Strehl bounded by the minimum $\sigma_{I}^2$ values of the HV$_{5/7}$ model shown in Fig.~\ref{fig:overhead_pairs}(b).  
Correspondingly, AO improved the range of $\eta$ from 40 dB $\geq -10\log (\eta)\geq 30$ dB to 37 dB $\geq -10\log (\eta)\geq 23$ dB.  

\begin{figure*}[t!]
	\includegraphics[width=0.90\textwidth]{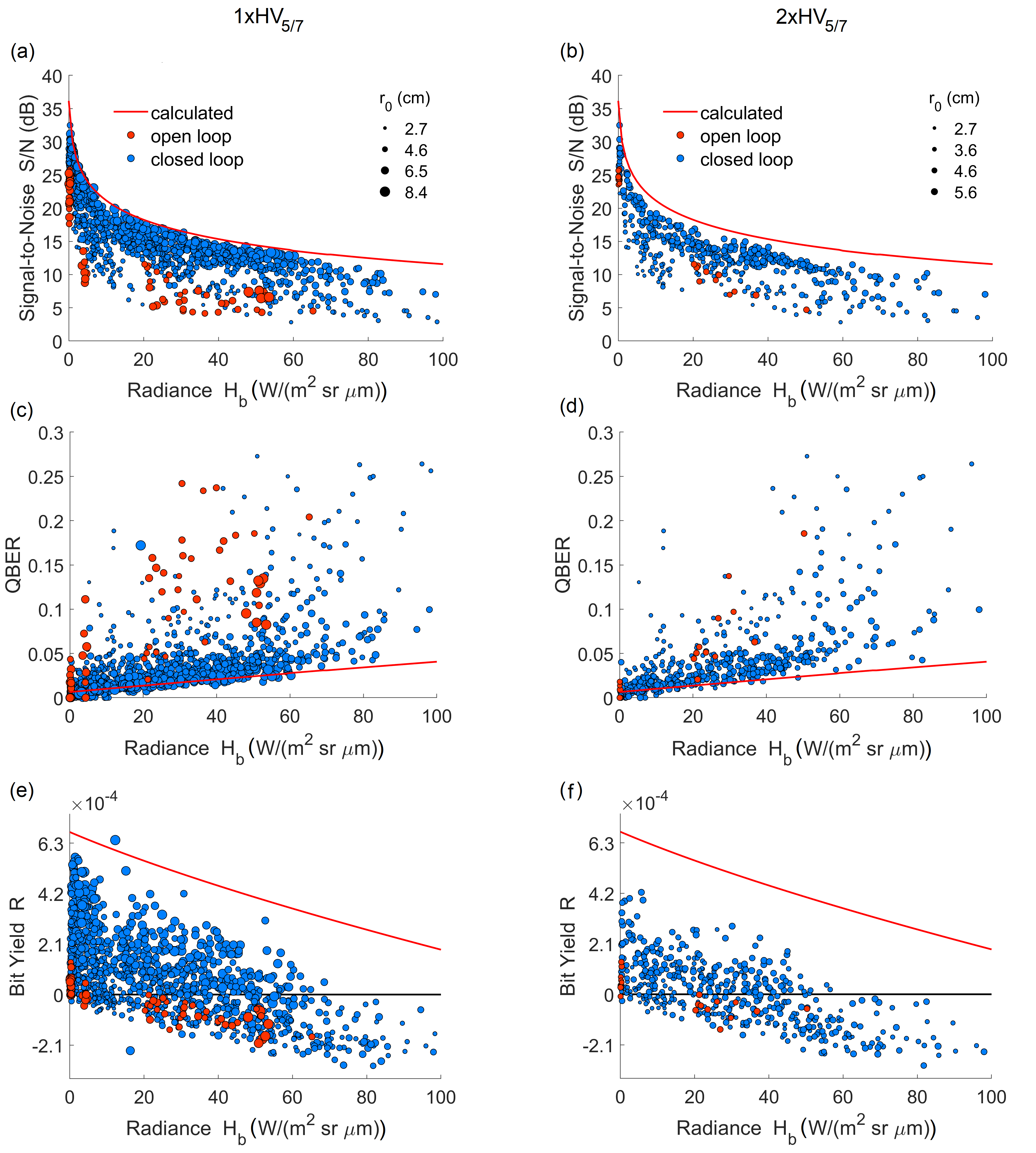}
\caption{\label{fig:results}
Plots of experimental results illustrating benefit of AO to daytime quantum channel performance including, (a) measured S/N probabilities, (b) measured QBER, and (c) calculated QKD bit yield vs. channel radiance $H_b$ under open-loop (red) and closed-loop (blue) conditions with $r_0$ represented by circle size. 
	}
\end{figure*}
Results summarizing the performance of the quantum channel as a function of channel radiance are presented in Fig.~\ref{fig:results}.  
We consider three statistical quantities relevant to QComm.  
First, we consider the probability a signal photon was detected relative to that for a noise photon, 
 \begin{equation}\label{eq:SNR}
\mathrm{S}/\mathrm{N} = Q_\mu / Y_0.
\end{equation}
Second, we consider the measured QBER, defined as the number of bit errors divided by the total number of measured qubits within a matched basis.  
Finally, based on measured quantum parameters, we calculate the bit yield probability for the vacuum-plus-decoy-state QKD protocol \cite{ma2005practical}.  
Results obtained with equivalent zenith angles falling on the 1$\times$HV$_{5/7}$ turbulence profile in Fig.~\ref{fig:overhead_pairs}(b) are shown in the left-hand column of Fig.~\ref{fig:results}.  
Those obtained with $\varepsilon_{\mathrm{zenith}}$ on the 2$\times$HV$_{5/7}$ profile are shown on the right.  
Open-loop cases are shown in red and closed-loop cases in blue.  
The measured values for $r_0$ are indicated by the circle sizes.  

The S/N probability ratio is plotted in dB for 1$\times$HV$_{5/7}$ and 2$\times$HV$_{5/7}$ turbulence strengths in Figs~\ref{fig:results}(a) and \ref{fig:results}(b), respectively.   
With higher-order AO, S/N probabilities range from a high of 1,775 when $H_b=0.35$ $\mathrm{W} / ( \mathrm{m}^2 \, \mathrm{sr} \, \mu \mathrm{m})$ to 10 at high radiance values near 80 $\mathrm{W} / ( \mathrm{m}^2 \, \mathrm{sr} \, \mu \mathrm{m})$.  
The solid red line, representing an upper bound, is calculated using the highest observed channel efficiency of $\eta$=23 dB with,
 \begin{equation}\label{eq:Y0}
Y_0 = N_b \, \eta_{\mathrm{rec}} \, \eta_{\mathrm{spec}} \, \eta_{\mathrm{det}} + 4 f_{\mathrm{dark}}  \, \Delta t,
\end{equation}
where $N_b$ was calculated using Eq.~\ref{eq:NUMBERDL}, $f_{\mathrm{dark}} = 190$-Hz is the measured APD dark count rate, and $\Delta t$ is the 1-ns detection window.  
Not surprisingly, experimental results lie closest to the calculated curve when $r_0$ values are largest.  
These data points represent the cases where the SHWFS is most able to resolve the spatial characteristics of turbulence.

Measured QBERs are plotted for 1$\times$HV$_{5/7}$ and 2$\times$HV$_{5/7}$ turbulence strengths in Figs~\ref{fig:results}(c) and \ref{fig:results}(d), respectively.   
Without AO, QBERs within 5$\%$ were only achieved at low daytime radiances. 
With AO, QBERs within 5$\%$ were achieved with $H_b$ as high as 80 $\mathrm{W} / ( \mathrm{m}^2 \, \mathrm{sr} \, \mu \mathrm{m})$.  
The solid red line shows the calculated QBER for the 23-dB channel according to,
 \begin{equation}\label{eq:QBER}
E_{\mu} = \dfrac{e_0 Y_0 + e_d (1 - e^{-\eta \mu})}{Y_0 + 1 - e^{-\eta \mu}},
\end{equation}
where the measured error rate and polarization crosstalk are $e_0 = 0.5$ and $e_d=0.005$, respectively.

Estimates for the QKD bit yield $R$ that could be achieved over the quantum channel were calculated from the measured signal- and decoy-state MPNs $\mu$ and $\nu$, signal- and decoy-state gains $Q_{\mu}$ and  $Q_{\nu}$, and signal- and decoy-state QBERs $E_{\mu}$ and $E_{\nu}$ assuming a constant efficiency of error correction $f(E_{\mu}) =1.22$.  
This calculation follows the formalism that is reviewed in Appendix~\ref{sec:appendixC} for the vacuum-plus-weak-decoy-state QKD protocol.  
Bit-yield probabilities are plotted for 1$\times$HV$_{5/7}$ and 2$\times$HV$_{5/7}$ turbulence strengths in Figs~\ref{fig:results}(e) and \ref{fig:results}(f), respectively.  
Without AO, positive bit yields were occasionally achieved with channel radiances below 4.3 $\mathrm{W} / ( \mathrm{m}^2 \, \mathrm{sr} \, \mu \mathrm{m})$, which is near a minimum for actual daytime sky radiances \cite{gruneisen2014adaptive, nordholt2002present, gruneisen2015modeling}.  
With AO, positive bit yields occur with channel radiances as high as 65 $\mathrm{W} / ( \mathrm{m}^2 \, \mathrm{sr} \, \mu \mathrm{m})$.  
The solid red line shows the bit yield $R$ calculated as a function of $H_b$ for the 23-dB channel.  
The solid black line at $R=0$ represents the level above which QKD should be possible.

\begin{figure*}[t!]
	\includegraphics[width=0.82\textwidth]{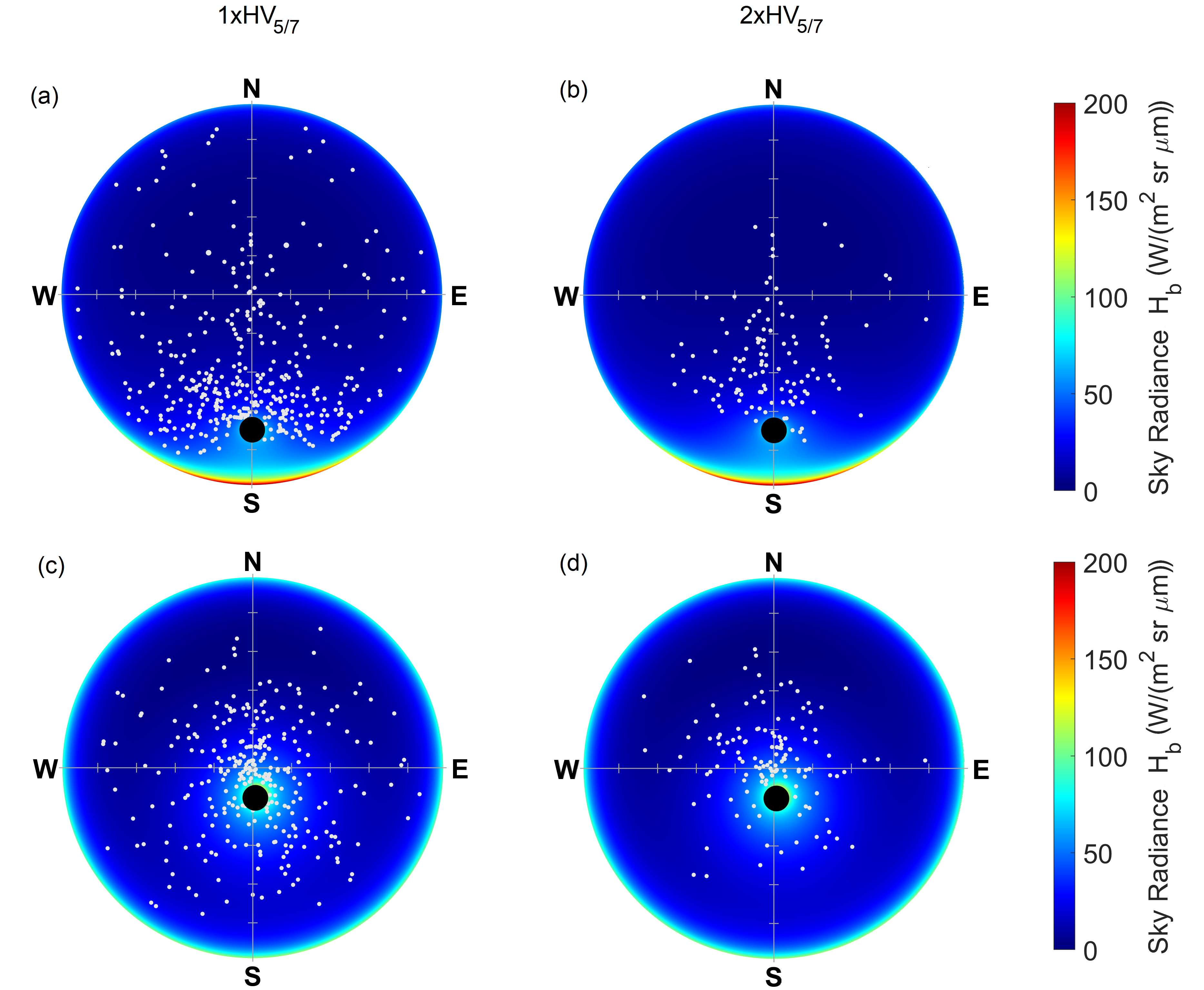}
\caption{\label{fig:hemi_plots_reduced}
Hemispherical plots, similar to Fig.~\ref{fig:hemi_plots} but retaining only sky angles for data sets that yield positive QKD bit-yield probabilities.   
	}
\end{figure*}
Not all equivalent zenith angles produced sufficiently high S/N or low QBER to yield a positive $R$. 
Figure~\ref{fig:hemi_plots} showed all simulated sky angles under which field-site data was taken whereas Fig.~\ref{fig:hemi_plots_reduced} retains only the simulated sky angles that yielded $R>0$.  
There are a sufficient number of simulated sky angles remaining to indicate viable quantum channels could be achieved over much of the daytime sky hemisphere in both 1$\times$HV$_{5/7}$ and 2$\times$HV$_{5/7}$ turbulence conditions.  
Comparing Figs.~\ref{fig:hemi_plots} and \ref{fig:hemi_plots_reduced}, it is not surprising that many data sets with simulated sky angles in close proximity to the sun fail to yield a positive $R$.  
Other data sets failed to yield a positive $R$ because, although they projected to a particular zenith angle on the HV$_{5/7}$ curves, they did so far from the curves and therefore represent much more demanding turbulence than the slant-path conditions for which the AO system was designed. 
In other words, in Fig.~\ref{fig:overhead_pairs}(b), multiple data sets project onto similar sky angles but some data sets have substantially higher $\sigma_I^2$ or substantially smaller $r_0$ than those describing slant path turbulence according to the HV$_{5/7}$ model.   
The absence of a data point in the sky hemisphere does not necessarily imply QComm would not be possible in that region of the sky.  
In most cases, it simply means there were no data for that particular combination of $r_0$, $\sigma_I^2$, and $H_b$.

\section{Discussion}

The field experiment was conducted within calendar year 2019.  
Success based on preliminary results was reported in a press release in May 2019 \cite{pressrelease}.  
Results and analysis were made publicly available and submitted for peer review in June 2020 \cite{gruneisen2020adaptive}.
Subsequently, there has been another report of free-space QComm using AO and claiming relevancy to space-Earth links but reported under ambiguous turbulence and background conditions \cite{cao2020long}.  
In that report, an iterative metric optimization technique was chosen over the SHWFS approach with the goal of accommodating the deep turbulence effects that occurred over an 11.5-km terrestrial path.  
The authors report that a 1-kHz-bandwidth 40-actuator-DM AO system compensated the 12 lowest-order Zernike modes and improved fiber-coupling efficiency by 6 to 8 dB.  
In contrast, our results were achieved with turbulence and channel radiance conditions representative of actual slant-path propagation.
In our system, A 2-kHz-frame-rate 130-Hz-bandwidth AO system compensated turbulence effects via zonal correction applied to the DM actuators that lie within the region defined by the circular pupil minus the central obscuration.   
This constituted approximately 82 actuators of the 12$\times$12 array that shape the continuous face sheet of the DM.
Previously we showed through simulation that a 200-Hz bandwidth 16x16 SHWFS-based AO system integrated with a 1-m telescope could improve fiber-coupling efficiencies in LEO downlinks by as much as 18 dB \cite{gruneisen2017modeling}.  

The choice of subaperture size relative to $r_0$ is an important consideration in AO design.  
In our experiment, the subaperture size of the 11$\times$11 SHWFS was 3.2 cm at the entrance pupil and our measured $r_0$ ranged from 2.0 cm to 8.5 cm in agreement with downlink turbulence values.  
For data sets where $r_0$ was comparable to or smaller than the subaperture size, under-resolved wavefront errors could have led to reduced AO performance and reduced $\eta_{\mathrm{FS}}$.  
Provided that the S/N ratio on the WFS camera can be maintained, increasing the number of SHWFS subapertures and reducing their size could improve results for these data sets.  
Were this particular AO system to be scaled to a larger aperture ground telescope for an actual satellite experiment where slewing is involved, increasing both the number of subapertures and the closed-loop bandwidth would benefit AO performance.

In our experiment, the receiver system operated at the D-L FOV.  
Results achieved with this configuration are generally relevant to systems where the receiver mode is restricted to a single spatial mode such as that of a single-mode optical fiber.  
For systems that are not constrained to the fundamental spatial mode, we have conducted a more detailed analysis that shows that one can further optimize quantum channel performance by choosing a FOV that accounts for residual wavefront errors that persist after AO compensation \cite{lanning2021quantum}.

\section{Conclusion}

In this article we report a QComm field experiment enabled by higher-order AO that is the first to demonstrate a pragmatic path toward a daytime QComm satellite downlink.  
The experiment was performed under carefully tuned downlink conditions.  
Namely, aperture coupling losses achieved via defocusing were representative of a 700-km downlink, turbulence conditions were representative of 1$\times$HV$_{5/7}$ turbulence strength over $0^{\circ} \leq \theta_z \leq 75^{\circ}$ and 2$\times$HV$_{5/7}$ turbulence strength over $0^{\circ} \leq \theta_z \leq 65^{\circ}$, and optical noise was representative of daytime on-sky conditions which are more than 1,000 times brighter than nighttime.  
This approach differs from those in earlier reports where transmitter beam divergence and aperture coupling losses were minimized and turbulence and background conditions were ambiguous.  
We have shown that a suitably designed higher-order AO system significantly reduces qubit losses when spatially filtering optical noise near the D-L.  
This permits the use of a relatively large 1-nm spectral filter, which is useful for integrating entangled photon sources of comparable bandwidth.  
AO can also enhance the efficiency of coupling into waveguide-based quantum components and networks and enhance the efficiency of Bell-state measurements for teleportation and entanglement swapping.  
The relevancy of these results can be extended to higher-altitude smaller-aperture satellites by increasing the spectral and temporal filtering of noise beyond that employed in this experiment.

\begin{acknowledgments}
The experimental system was developed and operated by The Boeing Company led by Mark Eickhoff.  Data analysis was performed by Leidos and led by Mark Harris.  The principal investigator for the effort was Mark Gruneisen, AFRL.  The authors gratefully acknowledge Margie Stewart, The Boeing Company, for aiding with scintillometer measurements, useful discussions with J. Frank Camacho, Leidos, and program management support from Capt. Keith Wyman and Valerie Knight, AFRL.  This work was supported by the Office of the Secretary of Defense (OSD) ARAP Defense Optical Channel Program (DOCP) and the Air Force Office of Scientific Research (AFOSR).  Approved for public release; distribution is unlimited. Public Affairs release approval AFRL-2021-1343.
\end{acknowledgments}

\bibliography{bibliography/bibliography}

\appendix
\section{Creating the spatial characteristics of slant-path turbulence at a terrestrial field site}\label{sec:appendixA}

The spatial nature of atmospheric turbulence is quantified through the spatial coherence and scintillation properties of the optical field.  
Spatial coherence is quantified through Fried's coherence length, $r_0$. 
Scintillation associated with the depth of the turbulent path is quantified through the log-intensity variance, or Rytov variance.   
Experimentally, scintillation is quantified via the scintillation index, $\sigma_I^2$, which is a measurable quantity that saturates with increased depth of turbulence.  

Figure~\ref{fig:overhead_pairs}(b) of the main text shows that the 1.6-km atmospheric path yielded ranges of $r_0$ and $\sigma_I^2$ that are comparable to those calculated for slant-path propagation through atmospheric turbulence over a large range of zenith angles.  
This appendix provides the theoretical basis for this experimental result and demonstrates the 1.6-km propagation path is what one would expect to be nearly optimum for introducing the spatial characteristics of slant-path turbulence.  

We begin by reviewing theoretical expressions for $r_0$, Rytov variance, and $\sigma_I^2$ as a function of the atmospheric structure parameter both for slant-path propagation and for horizontal propagation.  
We then proceed to calculate the horizontal path length that introduces $r_0$ and Rytov variance values comparable to those introduced by slant-path propagation.  
An important consequence of this analysis is that propagation over significantly longer distances actually misrepresents the effects of slant-path turbulence.    

The strength of turbulence through any propagation path is described by the structure parameter $C_n^2$.  In the case of slant-path propagation between space and Earth, the structure parameter takes the form of an altitude-dependent function, $C_n^2(h)$, where $h$ is the height above ground level.  The Hufnagel-Valley HV$_{5/7}$ model of $C_n^2(h)$ defines relevant conditions for many ground observational sites and can be expressed as \cite{sasiela2012electromagnetic},
 \begin{equation}\label{eq:CN2}
\begin{split}
C_{n}^{2}(h) &=0.00594 \, \Big( \frac{w}{27}\Big)^{2} ( 10^{-5} h)^{10} e^{-h/1000}\\
&+2.7 \times 10^{-16}    e^{-h/1500} +A e^{-h/100},
\end{split}
\end{equation}
where $C_n^2(h)$ is in m$^{-2/3}$, $h$ is in m, $w$ is a pseudo wind speed taken to be 21 m/s, and $A = 1.7 \times 10^{-14} $ m$^{-2/3}$.  
The strength of turbulence varies considerably over the course of a day and the HV$_{5/7}$ model can be scaled to approximate stronger turbulence through a multiplicative factor to $C_n^2(h)$.  
These scaled versions are designated by the multiplicative factor as 1$\times$HV$_{5/7}$, 2$\times$HV$_{5/7}$, 3$\times$HV$_{5/7}$, etc. 
Calculating the net effect of turbulence over any path requires integrating the structure parameter over the path.

\subsection{Fried Coherence Length}

The number and size of wavefront sensor subapertures required to resolve the spatial scale of turbulence is determined by $r_0$ which describes the transverse spatial scale of turbulence.  
For the case of long-distance propagation from a satellite to a telescope at ground level, the optical wavefront sampling the column of turbulence can be approximated as a plane-wave and $r_0$ can be expressed as \cite{tyson2015principles}, 
\begin{equation}\label{eq:R0SLANT}
\begin{split}
r_{0\, \mathrm{slant-path}} &=\Big(0.423 \,  k^{2} \sec(  \theta _{z} ) \\
&\times \int_0^a  C_{n} ^{2} (h) \, dh \Big)^{-3/5} , 
\end{split}
\end{equation}
where $k =2\pi/\lambda$, $\lambda$ is the optical wavelength, $\theta_z$ is the slant-path propagation angle relative to zenith and the upper limit of integration is the altitude of the satellite.  
For the field experiment conditions with diverging wavefronts over much shorter distances, the optical wavefront sampling the atmospheric path is approximated as a spherical wave and $r_0$ is given by \cite{tyson2015principles},
\begin{equation}\label{eq:R0HOR}
\begin{split}
r_{0\, \mathrm{horizontal}} =\Big(0.1602 \,  k^{2} \, C_{n} ^{2} (h_{0} ) \, L \Big)^{-3/5},   
\end{split}
\end{equation}
where $h_0$ is the height of propagation above ground level and $L$ is the horizontal propagation distance.

\subsection{Rytov Variance}

Over a sufficiently long propagation distance, turbulence-induced wavefront errors give rise to transverse intensity variations.  
In sufficiently deep turbulence, these intensity variations lead to intensity nulls that negatively impact the performance of a SHWFS-based AO system.  
The Rytov variance is the property of the atmosphere that gives rise to these intensity variations.  
For the case of long-distance propagation from a satellite to a telescope at ground level, the Rytov variance in the plane-wave approximation can be written as a function of zenith angle \cite{andrews2004field}:  
\begin{equation}\label{eq:RYTOVSLANT}
\begin{split}
\sigma _{1\, \mathrm{slant-path}}^{2} &=2.25 \,  k^{7/6} \, \sec^{11/6}( \theta _{z}) \\
&\times  \int_0^a C_{n} ^{2} (h ) \, h^{5/6} \, dh.
\end{split}
\end{equation}
For the field-site case, the Rytov variance in the spherical-wave approximation is given by \cite{andrews2004field},
\begin{equation}\label{eq:RYTOVHOR}
\begin{split}
\beta _{0\, \mathrm{horizontal}}^{2} =0.5 \,  k^{7/6} \, C_{n} ^{2} (h_{0} ) \, L^{11/6}.
\end{split}
\end{equation}

\subsection{Equivalent horizontal path lengths for $r_0$ and Rytov variance}

Setting Eqs.~\ref{eq:R0SLANT} and \ref{eq:R0HOR} equal and solving for $L$ yields the horizontal propagation distance needed to introduce the same $r_0$ that would be created by slant-path propagation at angle $\theta_z$:
\begin{equation}
\begin{split}
L_{r_0} &= \Bigg[  \dfrac{2.64  \int_{0}^{a} C_n^2 (h) \,dh}{C_n^2 (h_0)} \Bigg] \sec(\theta_z) \\
&=[377.1 \,\mathrm{m}] \sec(\theta_z),
\end{split}
\end{equation}
where the coefficient 377.1 m is found by numerical integration assuming $h_0=10$ m.  
Similarly, setting Eqs.~\ref{eq:RYTOVSLANT} and \ref{eq:RYTOVHOR} equal and solving for $L$ yields the horizontal propagation distance yielding the same Rytov variance created by slant-path propagation at angle $\theta_z$:
\begin{equation}
\begin{split}
L_{\mathrm{Rytov}} &= \Bigg[  \dfrac{4.5  \int_{0}^{a} C_n^2 (h) \, h^{5/6} \,dh}{C_n^2 (h_0)} \Bigg]^{6/11} \sec(\theta_z) \\
&= [682.1 \,\mathrm{m}] \sec(\theta_z),
\end{split}
\end{equation}
where the coefficient 682.1 m is found by numerical integration assuming $h_0=10$ m.  
Note that these results are independent of wavelength.  
Because stronger turbulence can occur in both slant-path and horizontal propagation, these results are also independent of the choice of turbulence strengths.  
The structure-parameter scaling coefficients apply equally to numerator and denominator and stronger turbulence does not affect the equivalent path calculation.

\subsection{Scintillation index}
The effects of scintillation are measured and characterized via the scintillation index $\sigma_I^2$ where, consistent with Ref.~\cite{andrews2004field}, the subscript “$I$” in $\sigma_I^2$ distinguishes this quantity from the plane-wave Rytov variance with subscript “1” in Eq.~\ref{eq:RYTOVSLANT}.  
For weak scintillation, the scintillation index is equal to the Rytov variance.  
As turbulence deepens, the Rytov variance increases but scintillation as measured by the scintillation index saturates.  
In the plane-wave approximation for slant-path turbulence, the scintillation index can be expressed as a function of the Rytov variance according to
\begin{equation}\label{eq:SCINTSLANT}
\begin{split}
\sigma _{I\, \mathrm{slant-path}}^{2} &= \exp \Bigg[ \frac{0.49 \, \sigma _{1}^{2} }{(1+1.11 \, \sigma _{1}^{12/5})^{7/6}} \\
&+\frac{0.51 \, \sigma _{1}^{2} }{(1+0.69 \, \sigma _{1}^{12/5})^{5/6}} \Bigg]-1.
\end{split}
\end{equation}
Correspondingly, in the spherical-wave approximation for horizontal propagation, 
\begin{equation}\label{eq:SCINTHOR}
\begin{split}
\sigma _{I\, \mathrm{horizontal}}^{2} &= \exp \Big[ \frac{0.49 \, \beta_{0}^{2} }{(1+0.56 \, \beta _{0}^{12/5})^{7/6}}\\
 &+\frac{0.51 \, \beta _{0}^{2} }{(1+0.69 \, \beta _{0}^{12/5})^{5/6}} \Big]-1.
\end{split}
\end{equation}
The maximum theoretical value for $\sigma_{I \, \mathrm{slant-path}}^2$ is approximately 1.24.  
The maximum calculated value for $\sigma_{I \, \mathrm{horizontal}}^2$ is approximately 1.69.  
This calculation assumes uniform $C_n^2$ over the path.  
Experimentally, the horizontal path is not always uniform and the maximum measured values were approximately 2.0.

\subsection{$[ \sigma_{I}^2, r_0]$ pairs in slant-path propagation}

\begin{figure}[t!]
	\includegraphics[width=1\columnwidth]{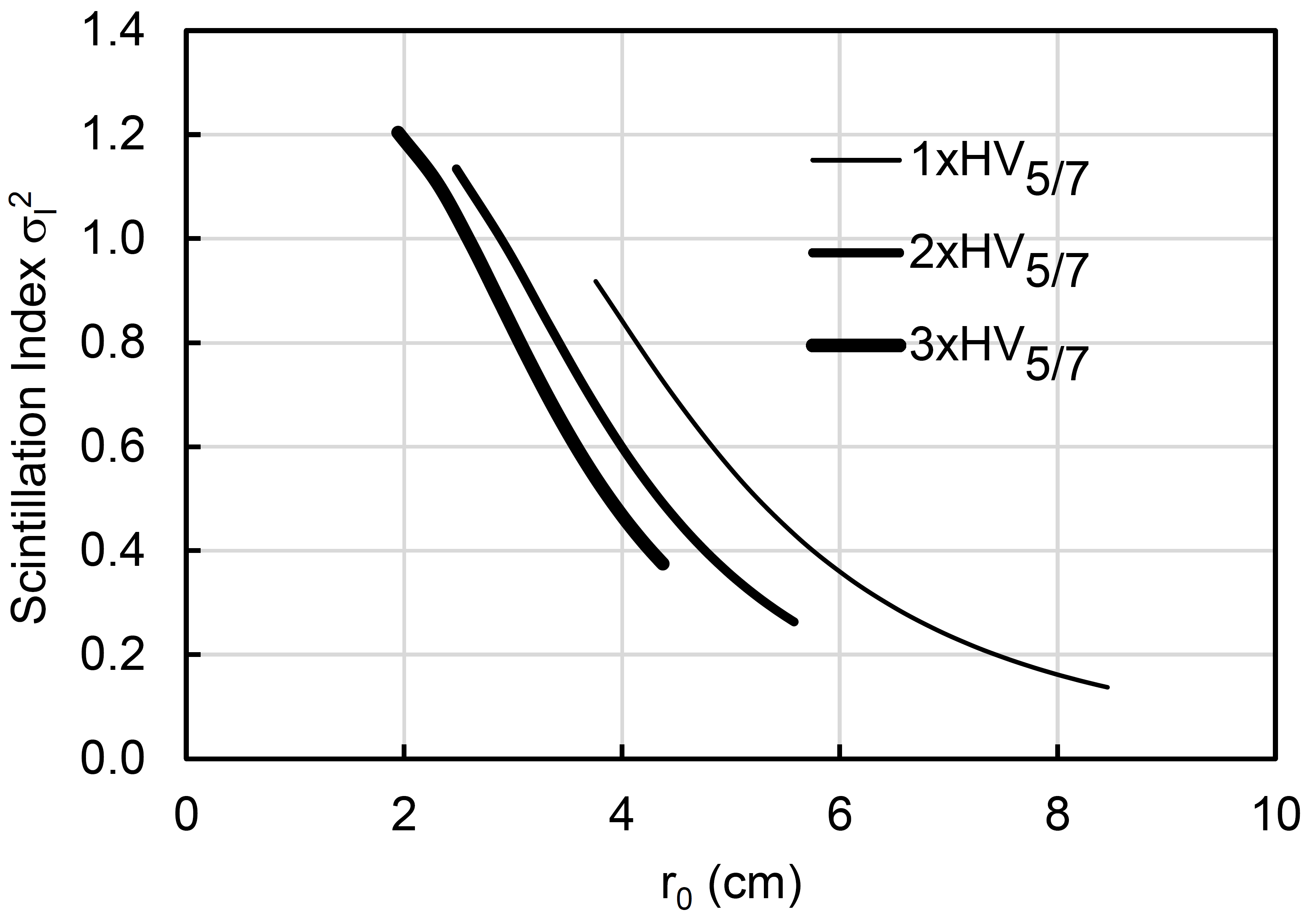}
\caption{\label{fig:scint_vs_r0}
Plot of $[ \sigma_I^2, r_0]$ pairs calculated for slant-path propagation with three strengths of turbulence and zenith angles ranging from 0$^{\circ}$ to 75$^{\circ}$.     
	}
\end{figure}

Figure~\ref{fig:scint_vs_r0} shows $[ \sigma_I^2, r_0]$ pairs calculated as a function of zenith angle from Eqs.~\ref{eq:R0SLANT} and \ref{eq:SCINTSLANT} for 780-nm wavelength and 1$\times$HV$_{5/7}$, 2$\times$HV$_{5/7}$, and 3$\times$HV$_{5/7}$ turbulence strengths and plotted over the range $0^{\circ} \leq \theta_z \leq 75^{\circ}$.  
Over this range of turbulence strengths, $r_0$ varies from 2.0 cm to 8.5 cm and $\sigma_I^2$ varies from 0.14 to 1.2.  
Turbulence at the field site varies considerably over the course of a day.  
This variability was utilized to create an experimental parameter space for $r_0$ that spanned the same range from 2.0- to 8.5-cm.  
The same natural variability yielded an experimental parameter space for $\sigma_I^2$ that spanned the range from $\sigma_I^2 = 0.14$, the minimum value occurring for 1$\times$HV$_{5/7}$ turbulence at $\theta_z = 0^{\circ}$, to $\sigma_I^2 \approx 0.9$, which corresponds to 2$\times$HV$_{5/7}$ turbulence at $\theta_z \approx 67^{\circ}$.  
In order to extend the range of $\sigma_I^2$, some of the data sets were acquired with a heat source placed under the beam path near the transmitter to increase scintillation.

\section{Temporal dynamics of slant-path turbulence relative to AO control-loop bandwidths}\label{sec:appendixB}

The temporal rate of change of turbulence is quantified through the Greenwood frequency $f_{\mathrm{G}}$ \cite{greenwood1977bandwidth}.  
For a static pointing angle, temporal fluctuations intrinsic to the atmospheric channel itself occur due to wind.  
When a ground telescopes tracks a moving satellite through turbulence, slewing leads to additional dynamics in the turbulence-induced wavefront error.  
Compensation of dynamic wavefront errors is most effective when the closed-loop control bandwidth of the AO system $f_c$ exceeds $f_{\mathrm{G}}$.  
However, $f_c$ does not denote a sharp cutoff frequency and degradation in AO compensation is typically graceful as $f_{\mathrm{G}}$ approaches and exceeds $f_c$.  

The 130-Hz closed-loop bandwidth of the AO system utilized in this field experiment was designed to accommodate the field-site turbulence where the maximum measured $f_{\mathrm{G}}$ was 62 Hz.  
Previously, we presented analyses quantifying the efficacy of 200- and 500-Hz bandwidth AO systems for a QKD protocol implemented from LEO.  
This appendix elucidates the relevancy of these three bandwidths to overhead LEO satellite passes.  
Greenwood frequencies are calculated with and without slewing effects and compared to AO bandwidths of interest. 
Circular LEO orbits and overhead passes are assumed.

\subsection{Greenwood frequencies versus zenith angle}

For propagation at a fixed angle, $f_{\mathrm{G}}$ is determined by altitude-dependent wind conditions and the strength of turbulence.  
The zenith-angle-dependent $f_{\mathrm{G}}$ can be estimated by the following equation \cite{tyson2015principles}: 
\begin{equation}\label{eq:FG}
\begin{split}
f_{\mathrm{G \, slant-path}} &= \Big[0.1022 \,  k^{2} \sec (  \theta _{z} ) \\
& \times \int_0^a  C_{n} ^{2} (h) \,  v^{5/3}(h) \, dh \Big]^{3/5}
\end{split}
\end{equation}
where $\theta_z$ is the zenith angle, $h$ is the height above ground level, and $C_n^2(h)$ is the altitude-dependent structure parameter given by Eq.~\ref{eq:CN2}.
It is common to assume the altitude-dependent Bufton wind profile \cite{sasiela2012electromagnetic}
\begin{equation}\label{eq:WIND}
\begin{split}
v(h)=v_{g}+30 \exp\Big[ -\Big(\frac{h-9400}{4800}\Big)^{2} \Big]
\end{split}
\end{equation}
where $v_g$ is the wind velocity near ground and is taken to be $v_g=5$ m/s. 

Tracking a moving satellite requires the ground telescope to slew at an angular rate that matches that of the satellite motion.  
The effect of slewing is to introduce an altitude-dependent contribution to the wind speed \cite{andrews2004field},
\begin{equation}\label{eq:SLEWWIND}
\begin{split}
v(h)= \omega _{s}h+v_{g}+30 \exp\Big[ -\Big(\frac{h-9400}{4800}\Big)^{2} \Big]
\end{split}
\end{equation}
where the angular rate of slewing $\omega_s$ depends on the satellite orbit.  
For the case of a satellite in a circular orbit passing directly overhead, $\omega_s$ can be written as \cite{andrews2004field},
 \begin{equation}\label{eq:SLEWRATE}
\begin{split}
\omega _{s}=\Big[ \frac{GM}{a^{2}(R+a)} \Big]^{1/2} \, \cos^{2}( \theta _z)
\end{split}
\end{equation}
where $G=6.673\times10^{-20}$ km$^3 \,$kg$^{-1} \,$s$^{-2}$ is the universal gravitational constant,  $M=5.97\times10^{24}$ kg is the mass of the Earth, $R=6,371$ km is the radius of the Earth, and $a$ is the satellite orbit altitude in km.  

\begin{figure}[t!]
	\includegraphics[width=.98\columnwidth]{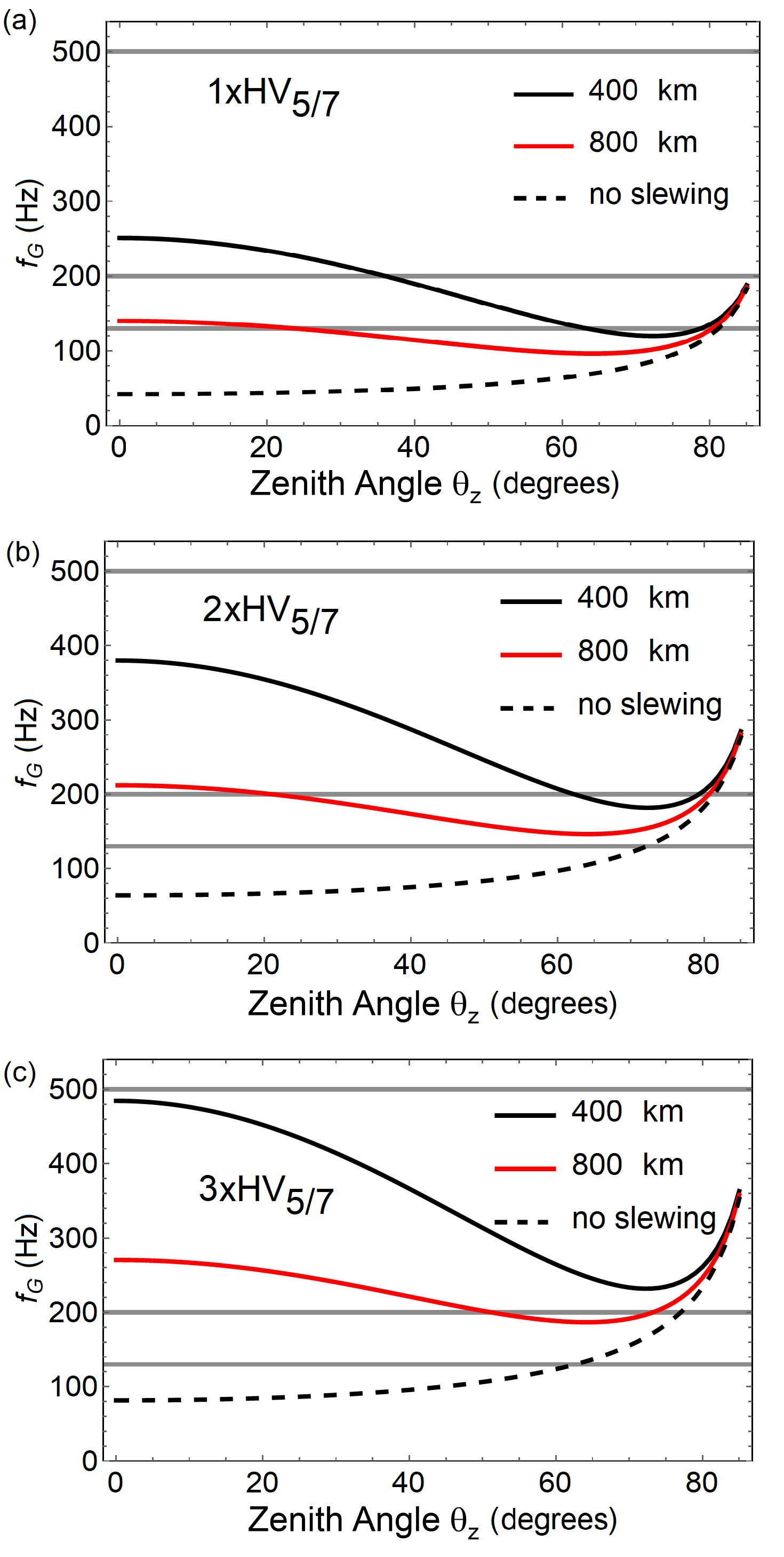}
\caption{\label{fig:fG_vs_zenith}
Plot of Greenwood frequency vs. zenith angle calculated for a) 1$\times$HV$_{5/7}$, b) 2$\times$HV$_{5/7}$, and c) 3$\times$HV$_{5/7}$ turbulence strengths.  Black and red solid lines are calculated for 400- and 800-km circular orbits, respectively.  The dashed lines are calculated without the effects of slewing.  Horizontal lines indicate the 130-, 200-, and 500-Hz AO loop bandwidths considered.
	}
\end{figure} 
Figure~\ref{fig:fG_vs_zenith} shows Greenwood frequencies calculated for $\lambda=780$-nm and plotted versus zenith angle.  
Greenwood frequencies calculated without slewing are shown as dashed lines.  
$f_{\mathrm{G}}$ values calculated for 400- and 800-km circular orbit altitudes are shown in black and red, respectively. 
 Without slewing, $f_{\mathrm{G}}$ increases with zenith angle due to the increased atmospheric path length.  
With slewing, $f_{\mathrm{G}}$ attains its maximum value at zenith where slew rates are highest.  
The AO control-loop bandwidth implemented in this field experiment and those considered in previous analyses are shown as horizontal lines at 130 Hz, 200 Hz, and 500 Hz \cite{gruneisen2016adaptive, gruneisen2017modeling, gruneisen2020adaptive}.  
Results calculated with 1$\times$HV$_{5/7}$, 2$\times$HV$_{5/7}$, and 3$\times$HV$_{5/7}$ turbulence strengths are plotted in Figs. \ref{fig:fG_vs_zenith}(a), \ref{fig:fG_vs_zenith}(b), and \ref{fig:fG_vs_zenith}(c), respectively.  

The 130-Hz control-loop bandwidth implemented in the field experiment is comparable to or greater than all stationary-angle $f_{\mathrm{G}}$ within a 120$^{\circ}$ cone about zenith.
Figure~\ref{fig:fG_vs_zenith}(a) shows that for an 800-km orbit and 1$\times$HV$_{5/7}$ turbulence, 130 Hz is also greater than all $f_{\mathrm{G}}$ within a 160$^{\circ}$ cone.
For cases with stronger turbulence or the lower orbit, $f_{\mathrm{G}}$ increases and typically exceeds the 130-Hz bandwidth.  

Previously, we presented a numerical simulation showing that a 200-Hz control-loop bandwidth AO system can be useful for implementations with 400-km and 800-km orbits even when $f_{\mathrm{G}}$ exceeds $f_c $ \cite{gruneisen2017modeling}.   
For the 800-km orbit case shown Figs.~\ref{fig:fG_vs_zenith}(a) and \ref{fig:fG_vs_zenith}(b), the 200-Hz bandwidth is comparable to, or exceeds, $f_{\mathrm{G}}$ for nearly all zenith angles.  
For the lower 400-km orbit altitude however, $f_{\mathrm{G}}$ can exceed 200 Hz at the smaller zenith angles. 
The 500-Hz control-loop bandwidth that was previously analyzed in Ref.~\cite{gruneisen2016adaptive} 
exceeds $f_{\mathrm{G}}$ for all cases considered in the overhead pass scenario.

\section{Quantum Channel Performance Measures}\label{sec:appendixC}

This appendix reviews definitions and equations associated with quantum measurements and measures of quantum channel performance.  
Measures of quantum channel performance include the S/N probability as well as the QBER and key bit yield associated with a decoy-state BB84 QKD protocol.  
The field experiment was conducted using a prepare-and-measure protocol implemented with weak-coherent pulses obeying Poissonian photon number statistics.  
Consistent with the vacuum-plus-decoy-state protocol \cite{ma2005practical}, the pulses launched from the transmitter included signal, decoy, and vacuum pulses of mean photon number $\mu$, $\nu$, and 0, respectively.  
Received pulses are detected via Geiger-mode APDs within a measurement window defined by a temporal interval $\Delta t$, spectral window $\Delta \lambda$, and solid-angle FOV $\Omega_{\mathrm{FOV}}$.

\subsection{Signal-to-Noise Probability Ratio} 

The probability a background detection event occurs due to background optical noise or detector dark counts is given by 
 \begin{equation}
Y_0 = N_b \, \eta_{\mathrm{rec}} \, \eta_{\mathrm{spec}} \, \eta_{\mathrm{det}} + 4 f_{\mathrm{dark}} \, \Delta t,
\end{equation}
where $N_b$ is the number of background photons in a detection window $\Delta t$.  
The efficiencies $\eta_{\mathrm{rec}}$, $\eta_{\mathrm{spec}}$, and $\eta_{\mathrm{det}}$ are the optical efficiencies of the receiver optics, spectral filter, and detectors, respectively, and $f_{\mathrm{dark}}$ is the dark count rate of the detectors.
For an optical receiver pointed to the sky, the parameter $N_b$ is proportional to the sky radiance $H_b$ according to,
\begin{equation}
N_b = \dfrac{H_b \, \Omega_{\mathrm{FOV}} \, \pi D^2_{\mathrm{R}} \lambda \, \Delta \lambda \, \Delta t }{4 h c},
\end{equation}
where $\lambda$ is the quantum channel wavelength, $\Delta \lambda$ is the spectral filter bandpass in $\mu$m, $\Delta t$ is the integration time for photon counting in seconds, $h$ is Planck's constant, and $c$ is the speed of light.

The probability of a detection event occurring within a measurement window is referred to as the signal-state gain $Q_{\mu}$.  
Because signal detection events are indistinguishable from background detection events, this is given by the sum of the probabilities for detecting a non-zero signal pulse and a noise photon
\begin{equation}\label{eq:c3}
Q_\mu = Y_0 + 1 - e^{-\eta \mu}.
\end{equation}
where $\mu$ is the mean photon number of the signal states and $\eta$ is the quantum channel efficiency including all contributions associated with transmitting and detecting signal pulses:
\begin{equation}
  \eta =\eta_\mathrm{geo} \,\eta_\mathrm{trans}  \,\eta_\mathrm{FS}  \, \eta_\mathrm{rec} \, \eta_\mathrm{spec} \, \eta_\mathrm{det},
\end{equation}
where $\eta_{\mathrm{geo}}$, $\eta_{\mathrm{trans}}$, and $\eta_{\mathrm{FS}}$ are the efficiencies associated with diffraction losses between transmitter and receiver apertures, atmospheric scattering and absorption, and losses due to the effects of atmospheric turbulence at the field stop, respectively.  
Note that in the downlink scenario, the effects of atmospheric turbulence on aperture-to-aperture coupling are negligible and therefore not included.

The signal-state gain $Q_{\mu}$ is measured experimentally during time intervals when signal pulses are launched. 
The background probability $Y_0$ is measured experimentally during intervals when no signal pulses are launched.  
The ratio of these two parameters gives the S/N probability which is a measure of quantum channel performance:
\begin{equation}
\mathrm{S}/\mathrm{N} = Q_{ \mu}/Y_{0}.
\end{equation}

\subsection{Quantum Key Distribution Measures}

For QKD protocols, the QBER $E_{\mu}$ is a useful measure of quantum channel performance.   
The QBER is defined within a matched basis as the probability an incorrect state is measured divided by the probability of any detection event occurring within a measurement window.  
$E_{\mu}$ was determined experimentally by comparing detected quantum states to the transmitted states.  
Theoretically, $E_{\mu}$ is calculated as a function of $H_b$ according to
 \begin{equation}\label{eq:c6}
E_{\mu} = \dfrac{e_0 Y_0 + e_d (1 - e^{-\eta \mu})}{Y_0 + 1 - e^{-\eta \mu}},
\end{equation}
where $e_0$ is the noise error rate and the polarization crosstalk $e_d$ is the probability a photon prepared in one linear polarization will be detected as the orthogonal polarization due to polarization crosstalk associated with imperfect optics.
The parameter $e_d$ is determined by measuring quantum detection events in each of the four polarization states while transmitting pulses in an orthogonal polarization.

An important measure of performance for a QKD protocol is the key-bit yield referring to the probability that a given signal photon results in a processed key bit.  
For the vacuum-plus-decoy-state BB84 QKD protocol, this can be expressed as \cite{ma2005practical}
 \begin{equation}
R \geq q\Big(-Q_{\mu }\,f(E_{ \mu}) \,H_{2}(E_{ \mu})+Q_{1}[1-H_{2}(e_{1})]\Big),
\end{equation}
where the protocol efficiency $q$ is 0.5, $f(E_{\mu})$ is the bidirectional error correction efficiency, $Q_1$ is the gain of the single-photon state given by
 \begin{equation}
Q_{1}= \frac{ \mu ^{2}e^{- \mu }}{ \mu  \nu - \nu ^{2}} \Big(Q_{ \nu }e^{ \nu }-Q_{ \mu }e^{ \mu } \frac{ \nu ^{2}}{ \mu ^{2}} - \frac{ \mu ^{2}-\nu ^{2}}{\mu ^{2}}Y_{0}\Big),
\end{equation}
and $e_1$ is the error rate of the single photon states, 
 \begin{equation}
e_{1}=  \frac{E_{ \nu }Q_{ \nu }e^{ \nu }-e_{ 0 }Y_{0} }{Y_{1} \nu },
\end{equation}
where $Q_{\nu}$ is the gain of the weak decoy state given by substituting $\nu$ for $\mu$ in Eq.~\ref{eq:c3} and $E_{\nu}$ is the weak-decoy-state QBER given by substituting $\nu$ for $\mu$ in Eq.~\ref{eq:c6}.   
The parameter $Y_1$ is the lower bound for the yield of the single-photon states,
 \begin{equation}
Y_{1}= \frac{ \mu}{ \mu  \nu - \nu ^{2}} \Big(Q_{ \nu }e^{ \nu }-Q_{ \mu }e^{ \mu } \frac{ \nu ^{2}}{ \mu ^{2}} - \frac{ \mu ^{2}-\nu ^{2}}{\mu ^{2}}Y_{0}\Big).
\end{equation}
Information leakage to a potential eavesdropper is quantified through the Shannon binary entropy function as a function of the single-photon error rate:
 \begin{equation}
H_{2}(e_{1})=-e_{1} \log_2(e_{1})-(1-e_{1}) \log_2(1-e_{1}).
\end{equation}
In this field experiment, $\mu$, $\nu$, $E_{\mu}$, $E_{\nu}$, $Q_{\mu}$, $Q_{\nu}$, $Y_0$, and $\eta$ are measured via quantum detection events.  
From these measured values, one can calculate the QKD bit yield R that could be achieved through post-processing of a raw key.  

\end{document}